\documentclass[english,aps,pra,two column,floatfix]{revtex4-1}
\usepackage[T1]{fontenc}
\usepackage[latin9]{inputenc}
\usepackage{color}
\usepackage{amsmath}
\usepackage{amssymb}
\usepackage{graphicx}
\usepackage{esint}

\makeatletter

\@ifundefined{textcolor}{}
{%
\definecolor{BLACK}{gray}{0}
\definecolor{WHITE}{gray}{1}
\definecolor{RED}{rgb}{1,0,0}
\definecolor{GREEN}{rgb}{0,1,0}
\definecolor{BLUE}{rgb}{0,0,1}
\definecolor{CYAN}{cmyk}{1,0,0,0}
\definecolor{MAGENTA}{cmyk}{0,1,0,0}
\definecolor{YELLOW}{cmyk}{0,0,1,0}
}

\makeatother

\usepackage{babel}
\begin{document}

\title{Probabilistic quantum phase-space simulation of Bell violations and
their dynamical evolution}

\author{L. Rosales-Z\'arate$^{1}$, B. Opanchuk$^{1}$, P. D. Drummond$^{1}$,
and M. D. Reid$^{1}$ }

\affiliation{$^{1}$Centre for Quantum and Optical Science, Swinburne University
of Technology, Melbourne 3122, Australia}
\begin{abstract}
Quantum simulations of Bell inequality violations are numerically
obtained using probabilistic phase space methods, namely the positive
P-representation. In this approach the moments of quantum observables
are evaluated as moments of variables that have values outside the
normal eigenvalue range. There is thus a parallel with quantum weak
measurements and weak values. Nevertheless, the representation is
exactly equivalent to quantum mechanics. A number of states violating
Bell inequalities are sampled, demonstrating that these quantum paradoxes
can be treated with probabilistic methods. We treat quantum dynamics
by simulating the time evolution of the Bell state formed via parametric
down-conversion, and discuss multi-mode generalizations.
\end{abstract}
\maketitle

\section{Introduction}

The importance and complexity of quantum dynamics has been emphasized
by many physicists, including Dirac and Feynman~\cite{Dirac1929,Feynman1982}.
We live in a dynamical, multi-mode universe described by quantum mechanics,
yet the equations involved quickly become too complex to solve. Quantum
simulations provide a means of carrying out such dynamical calculations.
In principle one may do this using either computational methods~\cite{Haake1979,Corney2006,Deuar2007,Alon2008,Gambetta2008,Trotzky2012},
or through a physical system that imitates the required properties~\cite{Feynman1982,Cirac2003,Jaksch2005,Buluta2009,Islam2011,Georgescu2014}.
However, while physical imitations can be useful, they are often unavailable
with the required parameter values. Universal quantum computers are
another possibility, and \textcolor{black}{realizations of up to $6$
qubits now exist}~\cite{Lanyon2011}, but this is too small for many
problems. 

Probabilistic quantum phase-space methods, which are potentially scalable,
are often the only practical route towards quantum simulation of large
systems~\cite{Opanchuk2013-early-universe}. Hence, this approach
is of great utility when simulating a multi-mode quantum system. This
is because number state methods are unable to handle exponential complexity.
Methods such as linearization fail when there are nonlinear effects.
However, these issues do not greatly increase the complexity of probabilistic
phase-space equations. The number of variables required increases
linearly, not exponentially, with the number of spatial modes. 

Such methods have been widely used to treat quantum dynamical problems
in quantum optics and atom optics. They have been used to model propagation
of radiation fields in superfluorescence~\cite{Haake1979}, reproducing
the observed delay statistics. They have been applied to dynamical
propagation of quantum solitons, where the observed entanglement and
squeezing was predicted to very high accuracy, including non-Markovian
thermal reservoirs~\cite{Drummond1993-solitons,Corney2006}. Simulations
of the quantum dynamics of critical points in parametric down-conversion
have been carried out~\cite{Dechoum2004}. Recently, very large three
dimensional systems of colliding Bose-Einstein condensates have been
treated~\cite{Deuar2007,Deuar2011,Krachmalnicoff2010}. These are
first-principles, multi-mode, quantum dynamical calculations of substantial
complexity, which show the potential of phase-space techniques.

An important question is whether these approaches can treat highly
non-classical states. We are especially interested in cases that violate
Bell inequalities. Here we investigate this issue, by demonstrating
that Bell states have a probabilistic mapping which can be computationally
sampled. Our motivation is to illustrate how these techniques work,
in a situation where the quantum behavior is readily understood. We
give a careful analysis of the different types of Bell inequalities,
their loopholes when present, and the techniques required to simulate
them, both for static and dynamic simulations. In this latter case
we focus on how the positive P-distribution can be used to perform
simulations of the time-evolution of the violation of Bell type inequalities.
The models used can be readily scaled to larger multi-mode treatments. 

We treat bipartite states using the Clauser-Horne~\cite{Clauser1974}
(CH) version of the Bell inequalities~\cite{Bell1964}, the Clauser-Horne-Shimony-Holt
(CHSH) inequalities~\cite{Clauser1969}, and the multi-particle generalizations
of the CH inequalities~\cite{Drummond1983,Reid2002} which are used
in photonic experiments and we called CHD inequality. Our main focus
here is on dynamical quantum simulations of these different Bell violations.
This requires an analysis of the different types of measurement strategy
used in these experiments. A summary of the static results for the
CHD Bell inequality is published elsewhere~\cite{Drummond2014-bell-sim}.
In Section~\ref{sec:Sampling-method-and} we explain in depth both
the computational strategy that allows probabilistic sampling of the
positive-P distribution and the methods used to obtain evaluate the
correlations for the static results of the CHD Bell inequality, as
well as the other inequalities. We also compare the relative sampling
errors obtained in the dynamic and static cases.

In the dynamical simulations, which illustrate these issues in specific
cases relevant to experiment, we simulate the simplest model of the
loophole-free violation of Bell type inequalities in parametric down
conversion experiments~\cite{Reid1986-violations,Shih1988,Ou1988,Kwiat1995},
using the positive P-representation to simulate different types of
Bell violation. For the CHSH Bell inequality we also take account
of the post-selection process that is often used in experiment, which
excludes null events. Surprisingly, these time-dependent quantum simulations
of Bell inequality violations have \emph{lower} sampling errors than
static cases. We also describe how to extend our simple model to complex
multi-mode systems. 

The phase-space distributions used in this paper are positive~\cite{Drummond1980,Hillery1984}.
They exist for all quantum states, and their statistical moments correspond
to correlations of the type observed in Bell violations. We focus
here on the usual photonic Bell state experiments which have been
experimentally studied, with an emphasis on recent parametric down-conversion
experiments~\cite{Reid1986-violations,Ou1988,Shih1988,Giustina2013,Christensen2013},
in order to give a simple model for dynamical simulations. Large-scale
Bell violations of multipartite systems have been treated elsewhere~\cite{Opanchuk2014-bell-sim}. 

In order to understand this approach, we emphasize some important
points. While stochastic, this technique is very different to conventional
path-integral Monte-Carlo methods. Path-integrals can give probabilistic
behavior for ground states or finite temperature steady states, but
they are not positive for real time dynamical evolution. Since quantum
systems have no objective reality until measured, phase-space quantum
simulation methods do not need to give dynamical evolution in terms
of classical paths. We therefore use a more general definition of
simulation. The probabilistic sampling that we employ gives quantum
physical moments $\langle\rangle_{qm}$ as equal to probabilistic
averages $\langle\rangle_{st}$ over variables that may not be eigenvalues
of observables. Thus, the mean value for the Pauli spin $\hat{\sigma}_{x}$
is given as $\langle\hat{\sigma}_{x}\rangle_{qm}=\langle\sigma_{x}\rangle_{st}$
where the stochastic variable $\sigma_{x}$ can assume values beyond
the eigenvalue spectrum, $1/2$ and $-1/2$. Such properties are closely
associated with the quantum notion of ``weak measurements'' and
``weak values''~\cite{Aharonov1988}. From the perspective of quantum
mechanics, this difference is not important, as long as one can predict
experimentally measurable correlations and operator averages.

The paper is organized as follows. First, in Section~\ref{sec:Bell-Inequalities}
we review and analyze several different Bell inequalities, in order
to examine which inequalities are most suitable for simulating a loophole-free
Bell test based on parametric down conversion~\textemdash{} which
does not generate a simple Bell state. In Section~\ref{sec:Positive-P-representation}
we discuss the positive P-representation, while Section~\ref{sec:Sampling-method-and}
treats static quantum simulation of a Bell state. In Section~\ref{sec:Dynamical-Simulations}
we demonstrate the quantum dynamical simulations of violations of
the CH and CHSH inequalities in parametric down conversion. Finally,
Section~\ref{sec:Conclusions} summarizes our conclusions.

\section{\label{sec:Bell-Inequalities}Bell Inequalities}

The assumption of a local hidden variable (LHV) theory (which assumes
local realism) leads to a constraint~\textemdash{} a Bell inequality~\textemdash{}
on the observable correlations of a physical system~\cite{Bell1964}.
These inequalities can be violated by quantum mechanics. In this paper
our goal is to provide a probabilistic quantum simulation of these
violations. There are many different Bell inequalities. Here, we describe
only the Bell inequalities to be considered in this paper. We focus
on three cases: The Clauser-Horne-Shimony-Holt (CHSH) inequality,
the Clauser-Horne (CH) inequality and the CH Bell inequality using
moments extended to $N$ photon pairs, which we will call the $N$-photon
CHD inequality.

\subsection{CHSH Inequality}

The CHSH Bell inequality was formulated to account for Bell's original
proposal, where two particles are emitted from a common source and
the measurement performed on each of the particles gives a binary
outcome. It allows rigorous tests of LHV theories in realistic experimental
scenarios where losses can be important.

In the case of two particles emitted from a common source, measurements
by spatially separated observers (usually called Alice and Bob) are
modeled in the LHV theory by taking random samples of a common set
of parameters (the hidden variables) symbolized by $\lambda$. Here,
$P(\lambda)$ denotes the probability distribution for the hidden
variables $\lambda$, which can be discrete or continuous. Measured
values are then functions of some local detector/ analyzer settings
$a$ and $b$ at each location, and the hidden parameters $\lambda$.
The value observed by Alice with detector settings $a$ is $A(a,\lambda)$,
and similarly $B(b,\lambda)$ is defined for Bob's value with detector
settings $b$. For binary outcomes, $A,B=\pm1$.

We now introduce the specific notation of $A(\theta,\lambda)$ and
$B(\phi,\lambda)$ for Alice and Bob with a variable analyzer setting
$\theta$ and $\phi$ respectively. In most experiments so far, $\theta$
and $\phi$ correspond to polarizer angles~\cite{Clauser1978,Giustina2013,Christensen2013,Aspect1982-EPRB}.
Here, the measurement event includes the selection of the measurement
settings $\theta,\phi$ at each site, done after the emission of the
particles. The measurement events are assumed to be space-like separated.
In a local hidden variable theory the correlations are thus obtained
from a probabilistic calculation of the form:
\begin{eqnarray}
E(A,B) & = & \int A(\theta,\lambda)B(\phi,\lambda)P(\lambda)d\lambda\,.\label{eq:LHV}
\end{eqnarray}

Clauser, Horne, Shimony and Holt obtained a version of Bell's inequality
known as the CHSH Bell inequality~\cite{Clauser1969,D'Espagnat1971},
which gives classical limits to the expected correlation for the above
experiment conducted by Alice and Bob, and is given by $E[A,B]-E[A,B^{\prime}]+E[A^{\prime},B]+E[A^{\prime},B^{\prime}]\leq2$,
where $A$, $A'$ and $B$, $B'$ are two sets of measurements made
by Alice and Bob. We rewrite this as 
\begin{eqnarray}
E(\theta,\phi)-E(\theta,\phi')+E(\theta',\phi)+E(\theta',\phi') & \leq & 2.\label{eq:CHSH-1}
\end{eqnarray}
Here $E(\theta,\phi)$ is the correlation, $\theta$ and $\theta^{\prime}$
are measurements at location $\mathcal{A}$ with two different analyzer
settings, while $\phi$ and $\phi^{\prime}$ are the corresponding
measurements at location $\mathcal{B}$. From Tsirelson\textquoteright s
theorem~\cite{Cirel'son1980}, it is known that $2\sqrt{2}$ is the
true upper bound for the left side of this inequality within quantum
mechanics, and that the operators giving this maximal value can always
be isomorphically mapped to the Pauli spin matrices. In the simulation
graphs, we will plot a normalized ratio and its LHV bound as: 
\begin{eqnarray}
S_{\mathrm{CHSH}} & = & \frac{1}{2}\left[E(\theta,\phi)-E(\theta,\phi')+E(\theta',\phi)+E(\theta',\phi')\right]\nonumber \\
 & \le & 1.
\end{eqnarray}

To violate the LHV bound quantum mechanically, one well-known route
proposed by Bell is for Alice and Bob to make Pauli spin measurements
on the spatially separated yet entangled particles created by the
spin-$1/2$ singlet state~\cite{Bell1964}. Alternatively, any measurements
with binary outcomes that have equivalent correlations in quantum
mechanics will suffice, for example, polarization entangled photon
pairs~\cite{Aspect1982-EPRB,Aspect1982-time-varying}. 

Such ``Bell'' states are described by the wave-function:
\begin{equation}
\left|\psi\right\rangle =\frac{1}{\sqrt{2}}\left(\left|1\right\rangle _{a}\left|-1\right\rangle _{b}-\left|-1\right\rangle _{a}\left|1\right\rangle _{b}\right),
\end{equation}
where here $|\pm1\rangle_{a/b}$ represents the eigenstate of Pauli
spin operator $S_{Z}$ for the system $a$/ $b$. The spin measurements
performed are as follows:

\begin{equation}
\begin{split}\hat{A} & =S_{z}\otimes I,\\
\hat{A}^{\prime} & =S_{x}\otimes I,\\
\hat{B} & =-\frac{1}{\sqrt{2}}\ I\otimes(S_{z}+S_{x}),\\
\hat{B}^{\prime} & =\frac{1}{\sqrt{2}}\ I\otimes(S_{z}-S_{x}).
\end{split}
\end{equation}
A calculation within quantum mechanics shows that, for the Bell state,
the Bell inequality is predicted to be violated:
\begin{eqnarray}
\langle\hat{A}\hat{B}\rangle-\langle\hat{A}\hat{B}^{\prime}\rangle+\langle\hat{A}^{\prime}\hat{B}\rangle+\langle\hat{A}^{\prime}\hat{B}^{\prime}\rangle & = & 2\sqrt{2}>2.
\end{eqnarray}

\subsection{Clauser-Horne Bell inequality}

The Clauser-Horne (CH) inequality was developed to test LHV theories
in situations relating to photons and polarizers where one of the
outcomes may not be detectable. The CH inequality has proved useful
for tests of LHV theories where the effect of detection inefficiencies
is significant. We discuss this inequality, because it will also prove
useful in establishing rigorous Bell tests of LHV theories where we
use parametric down conversion (PDC) as the photon pair source~\cite{Reid1986-violations,Ou1988,Shih1988,Giustina2013,Christensen2013}. 

This test is designed to detect nonclassical behavior in the prototypical
photonic Bell state, which is:
\begin{equation}
\vert1_{B}\rangle=\frac{1}{\sqrt{2}}\left(\hat{a}_{1}^{\dagger}\hat{b}_{1}^{\dagger}+\hat{a}_{2}^{\dagger}\hat{b}_{2}^{\dagger}\right)\vert0\rangle.
\end{equation}
Here the relative phase of the two terms is irrelevant~\textemdash{}
it just changes the polarizer settings~\textemdash{} and we choose
a positive sign for convenience. In the measurements, we consider
two modes of orthogonal polarization incident on the polarizer at
$\mathcal{A}$, and a second pair of orthogonal modes incident on
the polarizer at $\mathcal{B}$. The modes at $\mathcal{A}$ and $\mathcal{B}$
are denoted by boson operators $\hat{a}_{1}$ $\left(\hat{a}_{2}\right)$
and $\hat{b}_{1}$ $\left(\hat{b}_{2}\right)$ respectively. The two
polarizers are independently rotated to settings $\theta$ and $\phi$.
At each polarizer there are two possible output fields: the transmitted
and reflected modes. 

If a single photon is incident on one of the polarizers, there are
two outcomes which we label ``up'' and ``down'' depending on whether
the photon is detected in the transmitted or reflected mode respectively.
The transmitted and reflected modes at polarizer $\mathcal{A}$ are
defined by rotated operators
\begin{eqnarray}
\hat{c}_{+} & = & \hat{a}_{1}\cos\theta+\hat{a}_{2}\sin\theta\nonumber \\
\hat{c}_{-} & = & -\hat{a}_{1}\sin\theta+\hat{a}_{2}\cos\theta.\label{eq:cpm-1}
\end{eqnarray}
Similarly, at $\mathcal{B}$ the two outputs are defined by boson
operators:
\begin{eqnarray}
\hat{d}_{+} & = & \hat{b}_{1}\cos\phi+\hat{b}_{2}\sin\phi\nonumber \\
\hat{d}_{-} & = & -\hat{b}_{1}\sin\phi+\hat{b}_{2}\cos\phi.\label{eq:dpm-1}
\end{eqnarray}
An experimentalist can measure at each location whether a single photon
passes into the ``up'' or ``down'' mode at the polarizer. The
outcome of measurement is assigned the Pauli spin value $+1$ if ``up'',
and $-1$ if ``down''. In this way, one can establish the joint
probability $P_{++}^{AB}(\theta,\phi)$ for detecting one photon at
$\mathcal{A}$ ``up'' (i.e., in mode $c_{+}$) with setting $\theta$,
and one photon ``up'' (i.e., in mode $d_{+}$) at $\mathcal{B}$
with setting $\phi$. It is also possible to measure the marginal
probability $P_{+}^{A}(\theta)$ for detecting ``up'' at $\mathcal{A}$
with setting $\theta$. The marginal $P_{+}^{B}(\phi)$ is defined
similarly. 

We note that there can be a third outcome (apart from ``up'' and
``down'') at each detector. This is the null event where no photon
is detected. This outcome is usually given the value $A,B=0$. The
null outcome can occur either because of non-ideal detection efficiencies,
or because of the nature of the quantum state describing the incident
photon. For example, the state could be a vacuum state. For the parametric
down conversion (PDC) process, there will be a nonzero probability
for detecting zero and multiple photons at each ``up'' and ``down''
location.

Now we introduce the Clauser-Horne (CH) Bell inequality which is satisfied
for all LHV theories~\cite{Clauser1974,Clauser1978}:
\begin{eqnarray}
 &  & P_{++}^{AB}(\theta,\phi)-P_{++}^{AB}(\theta,\phi')+P_{++}^{AB}(\theta',\phi)+P_{++}^{AB}(\theta',\phi')\nonumber \\
 &  & \quad\leq P_{+}^{A}(\theta')+P_{+}^{B}(\phi).\label{eq:CH-1}
\end{eqnarray}
This type of Bell inequality has been studied extensively in the literature~\cite{Eberhard1993,Cabello2007,Brunner2007,Reid2002,Larsson1998,Larsson2001,Drummond1983}.
It is useful in establishing loophole-free violations of LHV theories
where the probability $\eta$ for detecting a photon incident on a
detector is reduced below unity~\cite{Giustina2013,Christensen2013}. 

For convenience in comparing graphs, we will define a normalized Bell
inequality and its LHV bound as:
\begin{eqnarray}
S_{\mathrm{CH}} & = & \frac{P_{++}^{AB}(\theta,\phi)-P_{++}^{AB}(\theta,\phi')+P_{++}^{AB}(\theta',\phi)+P_{++}^{AB}(\theta',\phi')}{P_{+}^{A}(\theta')+P_{+}^{B}(\phi)}\nonumber \\
 & \le & 1.
\end{eqnarray}
The CH inequality may be compared with the Clauser-Horne-Shimony-Holt
(CHSH) Bell inequality~(\ref{eq:CHSH-1}) which can be applied in
this case: 
\begin{equation}
E(\theta,\phi)-E(\theta,\phi')+E(\theta',\phi)+E(\theta',\phi')\leq2\,,\label{eq:chsh2}
\end{equation}
(or the version of the CHSH inequality introduced by Garg and Mermin~\cite{Garg1987}).
Here, 
\begin{equation}
E(\theta,\phi)=P_{++}^{AB}(\theta,\phi)+P_{--}^{AB}(\theta,\phi)-P_{-+}^{AB}(\theta,\phi)-P_{+-}^{AB}(\theta,\phi)
\end{equation}
 is the expected value of the product of the Pauli spin outcomes at
each detector. 

Low efficiencies $\eta$ will make violation of the inequalities~(\ref{eq:CH-1})
and~(\ref{eq:chsh2}) impossible, since the marginal probabilities
on the right side of the CH inequality scale as $\eta$, whereas the
joint probabilities on the left side of both inequalities scale as
$\eta^{2}$. One approach is to evaluate the inequalities over the
sub-ensemble of jointly detected counts only, but this introduces
assumptions that create the so-called ``detection loophole''~\cite{Pearle1970}.
Where detection efficiencies are fully taken into account, the CH
inequality generally becomes favorable compared to the CHSH Bell inequality,
because the ratio of left to right side is $\eta$ as compared to
$\eta^{2}$.

Experimentally, it has proved difficult to obtain a direct, loophole
free violation of these Bell inequalities, owing to detector inefficiencies~\cite{Cabello2007,Larsson1998,Larsson2001}.
With simulations, we can include efficiency factors or not, as we
choose; in this paper we simulate efficient detection. Nonetheless,
the above reasoning also motivates us to use the CH Bell inequality
for loophole-free tests where correlated joint null events are significant,
as in the PDC process. We will explain this in a later section.

\subsection{Bell inequalities with intensity moments}

For the experimental scenario described in the previous subsection,
one can reformulate the CHSH and CH Bell inequalities in terms of
intensity correlations~\cite{Drummond1983,Reid1986-violations}.
This is useful for our calculation of the inequalities using probabilistic
simulations, as it gives a reduced sampling error.

Consider the CHSH inequality. Since only one or zero photons is ever
detected at each detector, the spin product $E(\theta,\phi)$ can
be written as
\begin{equation}
E(\theta,\phi)=\langle(\hat{c}_{+}^{\dagger}\hat{c}_{+}-\hat{c}_{-}^{\dagger}\hat{c}_{-})(\hat{d}_{+}^{\dagger}\hat{d}_{+}-\hat{d}_{-}^{\dagger}\hat{d}_{-})\rangle.
\end{equation}
Similarly for the CH inequality, the probability $P_{++}^{AB}(\theta,\phi)$
becomes
\begin{equation}
P_{++}^{AB}(\theta,\phi)=\langle\hat{c}_{+}^{\dagger}\hat{c}_{+}\hat{d}_{+}^{\dagger}\hat{d}_{+}\rangle,
\end{equation}
and the marginal probability $P_{+}^{A}(\theta)$ is 
\begin{equation}
P_{+}^{A}(\theta)=\langle\hat{c}_{+}^{\dagger}\hat{c}_{+}\rangle.
\end{equation}
In this case, where the number of photons incident on each detector
is always less than or equal to $1$, the Bell inequality expressed
in terms of the intensity moments can be derived rigorously (since
always A, $B\leq1$, as explained in Section II.A) without any extra
assumptions that might introduce loopholes.

\subsection{$N$- photon CHD Bell inequalities}

Following Drummond~\cite{Drummond1983} and Reid~\cite{Reid2002},
we also consider Bell inequalities defined where one has more than
one photon incident on each detector, for the experimental scenario
described above. Consider the case where $N$ photons are incident
at each analyzer/detector. For example, we might consider a source
that emits correlated photon pairs in the state 
\begin{equation}
\vert N_{B}\rangle=\frac{\left(\hat{a}_{1}^{\dagger}\hat{b}_{1}^{\dagger}+\hat{a}_{2}^{\dagger}\hat{b}_{2}^{\dagger}\right)^{N}\vert0\rangle}{N!\left(N+1\right)^{1/2}}.\label{eq:cooperative-state-1}
\end{equation}
This state is a $2N$ photon generalization of the Bell singlet state,
which can be realized by a PDC~\cite{Reid2002,Howell2002}. One way
to treat this is to define higher spin outcomes~\cite{Mermin1980}.
However, it is also useful to extend the CH approach, and to redefine
the ``$+$'' event to be that where \emph{all} $N$ photons are
transmitted through the polarizer and, therefore, are detected in
mode $+$. 

We define the higher order intensity correlations
\begin{equation}
G^{IJ}(\theta,\phi,N)=\langle N\vert(\hat{c}_{+}^{\dagger})^{I}\hat{c}_{+}^{I}(\hat{d}_{+}^{\dagger})^{J}\hat{d}_{+}^{J}\vert N\rangle.\label{eq:Correlations}
\end{equation}
These are proportional to the probability of observing $I$ photons
of polarization $+$ at detector $A$ and $J$ photons of polarization
$+$ at detector $B$ (exactly for $I=J=N$ but only as a first approximation
otherwise). Then, we see that 
\begin{equation}
P_{++}^{AB}(\theta,\phi)=\kappa G^{NN}(\theta,\phi,N),
\end{equation}
where $\kappa$ is a proportionality constant. The marginal $P_{+}^{A}(\phi)$
can be defined by the correlation 
\begin{equation}
G^{IJ}(\theta,\infty,N)=\langle N|(\hat{c}_{+}^{\dagger})^{I}\hat{c}_{+}^{I}:\left(\hat{b}_{+}^{\dagger}\hat{b}_{+}+\hat{b}_{-}^{\dagger}\hat{b}_{-}\right)^{J}:|N\rangle.
\end{equation}
The ``$\infty$'' case and the corresponding correlation $G^{IJ}(\theta,\infty,N)$
stand for the same measurement as for $G^{IJ}(\theta,\phi,N)$, but
with no polarizer at the second detector. Since $N$ photons are incident
at each detector, the marginal is given by 
\begin{equation}
P_{+}^{A}(\theta)=\kappa G^{NN}(\theta,\infty,N),
\end{equation}
and similarly for $P_{+}^{B}(\phi)$. This of course is useful, because
for the CH inequality, the $\kappa$ constant will cancel, which means
we do not have to worry about calculating its value and also that
its smallness will not detract from a violation. We emphasize as above,
that for the case where there are strictly $N$ photons incident at
each location, the marginal can be calculated with the summation over
all possible outcomes at the first detector (the $\infty$ term).
Where there are also null events (zero detections) at one detector
but not the other~\textemdash{} as with inefficient detectors~\textemdash{}
the use of the joint correlation to describe the marginal will lead
to loopholes. 

For the state $|N_{B}\rangle$, which is a generalization of the Bell
singlet state, the correlation function will depend only on the magnitude
of the angle difference $\theta-\phi$. To simplify, we let this difference
be called $\varphi$. Also, for the states of interest, the marginals
are independent of $\theta$ or $\phi$, and there is symmetry so
that $P_{+}^{A}(\theta)=P_{+}^{B}(\phi)$. On looking at the CH inequality,
we then see that with the usual angle choice whereby $\theta$, $\phi$,
$\theta'$ $\phi'$ increase sequentially by $\varphi$~\cite{Bell1964},
the inequality has the form~\cite{Drummond1983}:
\begin{equation}
S_{\mathrm{CHD}}(\varphi)\equiv S_{N}^{N}(\varphi)=\frac{3g_{N}^{N}(\varphi)-g_{N}^{N}(3\varphi)}{2}\le0,\label{eq:BI_Cooperative}
\end{equation}
where
\begin{equation}
g_{N}^{J}(\varphi)=G^{JJ}(0,\varphi,N)/G^{JJ}(0,\infty,N).
\end{equation}
This expression generalizes the usual CH and Bell expressions to a
multi-particle form. We will call this Bell inequality the $N$-photon
CHD Bell inequality. The quantum mechanical prediction for $g_{N}^{J}$
for the state $|N\rangle$ has an especially simple form in case of
$J=N$: $g_{N}^{N}(\varphi)=\cos^{2N}\varphi$, which we use to test
our simulations.

\subsection{\label{sub:Bell-inequalities-PDC}Bell inequalities for parametric
down conversion}

One objective of this paper is to simulate the \emph{dynamical} generation
of the quantum state that violates a Bell inequality. In recent experiments,
violation of Bell inequalities has been achieved using parametric
down conversion (PDC). In this section, we therefore introduce a simple
dynamical model for this process, and consider which Bell inequalities
can be used to test LHV theories for PDC generated states.

Usually, in the experiments, the statistics are evaluated only on
a sub-ensemble of the measurement events. The sub-ensemble includes
only those events whereby a single photon is detected at each of the
spatially separated polarizers. In that case, it as though the Bell
state was incident on the two polarizers, and traditional approaches
(mentioned above) to testing Bell inequalities can be applied to test
local realism. For the case of the $N$-photon Bell inequalities,
one can similarly restrict to a $2N$-particle sub-ensemble. Such
projected measurements are typical of quantum optical tests of Bell
inequalities. This projection can be arranged in principle as a form
of state preparation (``heralding'')~\cite{D'Espagnat1971}. Also,
it can be arranged rigorously in the derivation of the CH-Bell inequality
by the suitable definition of the ``$+$'' measurement events~\cite{Reid2002},
a process we will describe below and will refer to as ``event selection''.
Both these forms of projection enable in principle a loophole free
test. However, more commonly, the projection is created by ``post-selection'',
after the detection of the photons. The post-selection procedure admits
a loophole, but has been necessary because of the null events created
by the inefficiency of detectors.

In this paper, we model the state generation by solving a Hamiltonian
that describes the PDC process~\cite{Reid1986-violations,Reid2002},
and consider testing Bell inequalities both with and without projection.
Here, we must address the null events that are created by the PDC
process. We show that it is possible to treat these events rigorously
without introducing loopholes, provided one uses the right Bell inequality. 

The experiments that test Bell's inequalities typically use either
an atomic cascade or parametric down-conversion, where the effective
Hamiltonian has the form:
\begin{equation}
\hat{H}=i\hbar\kappa E\left(\hat{a}_{1}^{\dagger}\hat{b}_{1}^{\dagger}-\hat{a}_{1}\hat{b}_{1}\right)+i\hbar\kappa E\left(\hat{a}_{2}^{\dagger}\hat{b}_{2}^{\dagger}-\hat{a}_{2}\hat{b}_{2}\right).\label{eq:HamPDC-1}
\end{equation}
where $\kappa$ denotes the strength of the parametric interaction,
and $E$ the strength of the pump field, and we will take $\hbar=1$.
Here $\hat{a}_{1}^{\dagger}$ creates a photon in site $A$ with polarization
1 (``$+$''), $\hat{a}_{2}^{\dagger}$ creates a photon in site
$\mathcal{A}$ with polarization 2 (``$-$''); similarly $\hat{b}_{1}^{\dagger}$
($\hat{b}_{2}^{\dagger}$) creates a photon in site $\mathcal{B}$
with polarization 1 (2). We suppose that $\hat{a}_{1,2}^{\dagger}$
($\hat{b}_{1,2}^{\dagger}$) creates a photon in the spatial modes
detected at site $\mathcal{A}$ ($\mathcal{B}$) with one of orthogonal
polarizations $1$ (2). We denote these modes by $A1$, $A2$, $B1$
and $B2$.

This model has just four relevant modes. By comparison, real experiments
are typically inhomogeneous and multi-mode in character~\cite{Shih1988,Ou1988,Kwiat1995,Howell2002},
since experimentalists usually employ traveling wave packets with
pulsed pump inputs to create the required spatial separation of detection
events. Such issues are readily treated with phase-space methods,
and full quantum field simulations have been treated elsewhere~\cite{Carter1987,Drummond1987,Werner1995}.
However, our purpose is not to exactly model an experiment, since
the details are different in every case. Instead, we wish to use this
simple model to understand the fundamental issues of probabilistic
sampling of quantum systems that violate a Bell inequality. A summary
of the required changes to generalize the simulations is given in
Section~\ref{sub:Extended-simulations}.

The four-mode Hamiltonian generates a correlated squeezed state, with
the generic form for $\kappa Et\ll1$ of:
\begin{equation}
\begin{split}\left|\psi\right\rangle = & \exp\left(-i\hat{H}t\right)\left|0\right\rangle \\
= & \left|0\right\rangle +\kappa Et\Bigl(\left|1\right\rangle _{A1}\left|0\right\rangle _{A2}\left|1\right\rangle _{B1}\left|0\right\rangle _{B2}\\
 & +\left|0\right\rangle _{A1}\left|1\right\rangle _{A2}\left|0\right\rangle _{B1}\left|1\right\rangle _{B2}\Bigr)+O\left(\kappa Et\right)^{2}.
\end{split}
\end{equation}
This does not generate just the Bell state $|1_{B}\rangle$. Instead,
for $\kappa Et\ll1$ it generates a linear superposition of the Bell
state and a correlated vacuum state $|0\rangle\equiv|0\rangle_{A1}|0\rangle_{A2}|0\rangle_{B1}|0\rangle_{B2}$.
Then, the generated quantum state has the form:
\begin{equation}
\left|\psi\right\rangle \approx\vert0\rangle+c\vert1_{B}\rangle,
\end{equation}
where $c=\sqrt{2}\kappa Et$.

Earlier, we considered the Clauser-Horne (CH) Bell-type inequality
\begin{eqnarray}
 &  & P_{++}^{AB}(\theta,\phi)-P_{++}^{AB}(\theta,\phi')+P_{++}^{AB}(\theta',\phi)+P_{++}^{AB}(\theta',\phi')\nonumber \\
 &  & \quad\leq P_{+}^{A}(\theta')+P_{+}^{B}(\phi).\label{eq:CH}
\end{eqnarray}
Here, we can define the ``$+$'' event at each detector to be where
$N$ photons are detected at the ``$+$'' polarized mode \emph{and}
a total of $N$ photons are detected in total at the ``$+$'' and
``$-$'' modes~\cite{Reid2002}. The CH Bell inequality is effective
for loophole-free tests in the presence of correlated joint null events,
which are significant in the PDC process due to the presence of the
correlated vacuum state $|0\rangle$, the leading term for low $\kappa Et$.
These null events will lead to a reduction in the absolute value of
the joint probabilities (such as $P_{++}^{AB}(\theta,\phi)$), which
substantially reduces the violation of the CHSH-type inequalities,
unless heralding or some other strategy can be utilized. For the case
of the ideal parametric amplifier, the joint null events are correlated.
As a result, because the CH inequality is normalized by the marginals
on the right-side, these vacuum events will have no impact on the
violation of the CH inequality~(\ref{eq:CH}). Also, for PDC, the
event of a total of $N$ particles being detected at one polarizer
is correlated with a total of $N$ particles detected at the other
polarizer. Therefore, this strategy is useful for projecting out the
$N$-particle Bell state. We call this strategy ``event selection''
and note it is useful in providing loophole-free tests for the $N$-photon
CHD Bell inequalities (using PDC).

In the following sections we explain how to probabilistically simulate
the dynamics of the generation of Bell violations for PDC experiments,
by using the simple model~(\ref{eq:HamPDC-1}).

\section{\label{sec:Positive-P-representation}Positive P-representation}

There are a number of different positive phase-space representations.
For bosonic systems, the most general class known extends the $s$-ordered
representations of Cahill and Glauber~\cite{Cahill1969} to include
the set of all Gaussian operator bases~\cite{Corney2003,Corney2004},
defined over nonclassical phase-space coordinates.

\subsection{Definition and existence properties}

The most well-known of these nonclassical phase-space methods is the
positive P-representation~\cite{Drummond1980}, which generalizes
the Glauber-Sudarshan P-representation~\cite{Glauber1963-states,Sudarshan1963}
to all quantum states. 

For $M$ bosonic modes, this is a non-unique expansion of an arbitrary
density matrix $\hat{\rho}$ in coherent state projectors:
\begin{equation}
\hat{\rho}=\int P(\vec{\alpha},\vec{\alpha}^{+})\hat{\Lambda}(\vec{\alpha},\vec{\alpha}^{+})d^{2M}\vec{\alpha}d^{2M}\vec{\alpha}^{+},
\end{equation}
where $\hat{\Lambda}$ is a coherent state projection operator, defined
as:
\begin{equation}
\hat{\Lambda}(\vec{\alpha},\vec{\alpha}^{+})=\frac{\left|\vec{\alpha}\right\rangle \langle\left(\vec{\alpha}^{+}\right)^{*}\vert}{\langle\left(\vec{\alpha}^{+}\right)^{*}\vert\vec{\alpha}\rangle}.
\end{equation}
Here $\left|\vec{\alpha}\right\rangle =\left|\alpha_{1}.\ldots\alpha_{M}\right\rangle $
is a multi-mode coherent state of a bosonic field, which is an eigenstate
of the corresponding operators $\left(\hat{a}_{1},\ldots\hat{a}_{M}\right)$.
The probability function $P(\vec{\alpha},\vec{\alpha}^{+})$ is defined
on an\textbf{ }enlarged, nonclassical phase-space, which allows positive
probabilities for all quantum states. 

This representation maps bosonic quantum states into $4M$ real coordinates:\textbf{
\begin{eqnarray}
\vec{\alpha} & = & \vec{p}+i\vec{x}\nonumber \\
\vec{\alpha}^{+} & = & \vec{p}^{+}+i\vec{x}^{+}\,,
\end{eqnarray}
}which is double the dimension of the corresponding classical phase-space.
This method leads to exact probabilistic mappings between quantum
mechanics and a classical-like phase-space description, even for low
occupation numbers. It is often advantageous to perform a variable
change to sum and difference variables:
\begin{eqnarray}
\vec{\nu} & = & \left(\vec{\alpha}-\left(\vec{\alpha}^{+}\right)^{*}\right)/2,\nonumber \\
\vec{\mu} & = & \left(\vec{\alpha}+\left(\vec{\alpha}^{+}\right)^{*}\right)/2.\label{eq:P-variable-change}
\end{eqnarray}

A general probabilistic construction using these variables, which
is non-unique but always exists, is~\cite{Drummond1980}:
\begin{equation}
\begin{split}P(\vec{\alpha},\vec{\alpha}^{+})= & \frac{\left\langle \vec{\mu}\right|\widehat{\rho}\left|\vec{\mu}\right\rangle }{\left(2\pi\right)^{2M}}e^{-\left|\vec{\nu}\right|^{2}}\end{split}
.\label{eq:P-from-rho}
\end{equation}
We will use this distribution for the static sampling calculations.
However, our dynamical sampling calculations do not employ this form,
but rather use a dynamically generated distribution, obtained from
solving coupled stochastic equations. 

In all cases, with the positive P-distribution, the expectation of
any normally ordered observable $\hat{O}\equiv O(\hat{a}_{1}^{\dagger},\hat{a}_{1},\ldots)$
is:
\begin{equation}
\left\langle \hat{O}\right\rangle =\int O(\alpha_{1}^{+},\alpha_{1},\ldots)P(\vec{\alpha},\vec{\alpha}^{+})d^{2M}\vec{\alpha}\, d^{2M}\vec{\alpha}^{+}.\label{+Pcorrels}
\end{equation}

For Bell state measurements, the effects of a polarizer are simply
obtained on taking linear combinations of mode amplitudes, just as
in classical theory or with quantum operators~\cite{Reid1986-violations}.
If we represent the input operators $\left(\hat{a}_{1},\hat{a}_{2},\hat{b}_{1},\hat{b}_{2}\right)$
by complex variables $\left(\alpha_{1},\alpha_{2},\beta_{1},\beta_{2}\right)$,
the transmitted and reflected modes at polarizer $A$ are defined
by rotated complex phase-space variables:
\begin{eqnarray}
\gamma_{+} & = & \alpha_{1}\cos\theta+\alpha_{2}\sin\theta\nonumber \\
\gamma_{-} & = & -\alpha_{1}\sin\theta+\alpha_{2}\cos\theta,\label{eq:cpm-gamma}
\end{eqnarray}
for the ``up'' and ``down'' modes respectively. Similarly, at
$B$ the outputs are defined by boson operators:
\begin{eqnarray}
\delta_{+} & = & \beta_{1}\cos\theta+\beta_{2}\sin\theta\nonumber \\
\delta_{-} & = & -\beta_{1}\sin\theta+\beta_{2}\cos\theta,\label{eq:cpm-delta}
\end{eqnarray}

The corresponding hermitian conjugate terms are represented by replacing
$\alpha,\beta,\gamma,\delta$ by independent complex variables $\alpha^{+},\beta^{+},\gamma^{+},\delta^{+}$.
The advantage is that this can represent entangled states: a positive
P-function always exists for any density matrix. In particular, it
exists for the photonic Bell state.

Since this is always probabilistic, there is a great similarity between
the hidden variable theory~(\ref{eq:LHV}) of Bell, and the positive-P
formula~(\ref{+Pcorrels}) for quantum correlations, from setting
$\lambda=\left(\vec{\alpha},\vec{\alpha}^{+}\right)$. However, while
the hidden variable theory obeys Bell's theorem and hence cannot be
equivalent to quantum theory, the positive-P theory is fully equivalent
to quantum mechanics, and therefore \emph{can} violate Bell's inequalities. 

The reason for the difference is due to the different quantities calculated
in the correlations~\cite{Reid1986-violations}. The fundamental
observables in Bell's case, of the form $X(\lambda)$, are defined
as being equal to actual observed real numbers, that is, $(0,1,\ldots)$
for photon counts. The corresponding observables in the positive-P
case, of form $n\left(\vec{\alpha},\vec{\alpha}^{+}\right)$, are
complex numbers whose mean values and correlations correspond to observable
means and correlations. Given these unrestricted numbers, the proof
of the Bell inequality is no longer applicable.

This difference allows the positive P-distribution to be exactly equivalent
to quantum mechanics, even though it appears in other respects just
like a probabilistic hidden variable theory. As a result, this approach
is well-suited to carrying out probabilistic quantum simulations.
This property of having quasi-observable parameters different to eigenvalues,
is also shared by weak quantum measurement strategies~\cite{Aharonov1988}.

\subsection{Cooperative Bell state distribution}

The four-mode state~(\ref{eq:cooperative-state-1}) has the corresponding
positive P-distribution~\cite{Drummond1983}:

\begin{widetext}

\begin{equation}
P(\vec{\alpha},\vec{\alpha}^{+})=\left\{ \frac{\left|\left(\alpha_{1}^{+}+\alpha_{1}^{*}\right)\left(\beta_{1}^{+}+\beta_{1}^{*}\right)+\left(\alpha_{2}^{+}+\alpha_{2}^{*}\right)\left(\beta_{2}^{+}+\beta_{2}^{*}\right)\right|^{2N}}{\left(2\pi\right)^{8}\left(N+1\right)\left(N!\right)^{2}2^{4N}}\right\} \exp\left(-\frac{\left|\vec{\alpha}\right|^{2}+\left|\vec{\alpha}^{+}\right|^{2}}{2}\right).\label{eq:cooperative-P}
\end{equation}

\end{widetext}For the positive-P function in the form of~(\ref{eq:P-from-rho}),
we perform the variable change given in Eqn.~(\ref{eq:P-variable-change}),
which has a Jacobian $2^{2M}$. For this four-mode distribution of
interest, we additionally introduce four complex vector functions
that describe the phase-space variables corresponding to measurements
at A and B respectively, giving a total of $16$ real dimensions.
These are:
\begin{eqnarray}
\vec{A} & = & \left[\mu_{A1},\mu_{A2}\right],\quad\vec{B}=\left[\mu_{B1},\mu_{B2}\right],\nonumber \\
\delta\vec{A} & = & \left[\nu_{A1},\nu_{A2}\right],\quad\delta\vec{B}=\left[\nu_{B1},\nu_{B2}\right].
\end{eqnarray}
Then the positive P-distribution~(\ref{eq:cooperative-P}) can be
written in the form of:
\begin{equation}
P(\vec{A},\vec{B},\delta\vec{A},\delta\vec{B})=P(\vec{A},\vec{B})G(\delta\vec{A})G(\delta\vec{B})\,.
\end{equation}
Here we have introduced an auxiliary distribution of:
\begin{equation}
P(\vec{A},\vec{B})=\left(\frac{\left|\vec{A}\cdot\vec{B}\right|^{2N}}{\pi^{4}\left(N+1\right)\left(N!\right)^{2}}\right)e^{-\left|\vec{A}\right|^{2}-\left|\vec{B}\right|^{2}},
\end{equation}
together with normal distributions:
\begin{equation}
G(\delta\vec{A})=\frac{1}{\pi^{2}}e^{-\left|\delta\vec{A}\right|^{2}}.
\end{equation}

\section{Sampling method and static Bell violations\label{sec:Sampling-method-and}}

In order to use probabilistic methods for the static distributions
given in the previous section, it is necessary to have a computational
algorithm that generates probabilistic samples. 

We can sample $\delta\vec{A}$, $\delta\vec{B}$ as four dimensional
Gaussian variates in real space, with a real variance of $\sigma^{2}=1/2$
in each real coordinate. Here we note that:
\begin{eqnarray}
\int e^{-\left|\vec{A}\right|^{2}}d^{2}\vec{A} & = & \left[\pi\int e^{-R}dR\right]^{2}=\pi^{2}.
\end{eqnarray}
Hence, the 8 real difference coordinates can all be exactly sampled
without rejection.

\subsection{Von Neumann Rejection Algorithm}

To sample $P(\vec{A},\vec{B})$ over the $8$ remaining real variables,
we choose the well-known von Neumann rejection algorithm, which is
an easily implemented technique. The algorithm used here relies on
sampling with a distribution proportional to a positive, normalizable
function $\tilde{F}(\vec{A},\vec{B})$ that is always larger than
the target distribution. Once sampled, the numbers generated are randomly
accepted or rejected in proportion to $P/\tilde{F}$, to obtain samples
with the required distribution. Since it is clear that:
\begin{equation}
\left|\vec{A},\vec{B}\right|\le\left|\vec{A}\right|\left|\vec{B}\right|,
\end{equation}
 we can choose the computational function according to:
\begin{equation}
P\left(\vec{A},\vec{B}\right)\le\tilde{F}\left(\vec{A},\vec{B}\right)=F\left(\vec{A}\right)F\left(\vec{B}\right),
\end{equation}
where $F\left(\vec{A}\right)$ has the following structure:
\begin{equation}
F\left(\vec{A}\right)=\frac{|\vec{A}|^{2N}}{\pi^{2}\sqrt{N+1}N!}\exp\left(-|\vec{A}|^{2}\right).
\end{equation}

Hence, we can use the rejection method described above. We note that
it might be feasible to use hyper-spherical coordinates and sample
this without rejection, but as the algorithm described here works
well, we did not attempt this refinement.

The function $F\left(\vec{A}\right)$ has to be normalized to establish
the acceptance/rejection ratio. In order to do that, we notice that:
\begin{eqnarray}
\int|\vec{A}|^{2N}\exp\left(-|\vec{A}|^{2}\right)d^{k}\vec{A} & = & S_{k-1}(1)\int_{0}^{\infty}r^{2N+k-1}e^{-r^{2}}dr\nonumber \\
 & = & \frac{1}{2}\Gamma(N+k/2)S_{k-1}(1),
\end{eqnarray}
where $k$ is the number of components in $\vec{A}$. In this case
$\vec{A}$ contains two complex numbers, so $k=4$, and $S_{k-1}(r)$
is the surface area of a $k$-dimensional ball:
\begin{eqnarray}
S_{k-1}(r) & = & \frac{2\pi^{k/2}r^{k-1}}{\Gamma(k/2)}\,.
\end{eqnarray}

Therefore, the normalization gives:
\begin{eqnarray}
\mathcal{N} & = & \int F(\vec{A})d^{4}\vec{A}\nonumber \\
 & = & \frac{\Gamma(N+2)}{2\pi^{2}\sqrt{N+1}N!}\times\frac{2\pi^{2}}{\Gamma(2)}.\nonumber \\
 & = & \sqrt{N+1}
\end{eqnarray}
and $F(\vec{A})=\mathcal{N}\tilde{P}(\vec{A})$, where $\tilde{P}$
is a probability distribution:
\begin{equation}
\tilde{P}(\vec{A})=\frac{|\vec{A}|^{2N}}{\pi^{2}(N+1)!}\exp\left(-|\vec{A}|^{2}\right),
\end{equation}
which is a combination of a lambda-distribution of the vector length
and a uniform distribution of its direction and therefore can be sampled
exactly.

It can be represented as a product of two independent distributions~\cite{Gupta1997}:
\begin{equation}
\tilde{P}(r,\vec{n})=S_{k-1}(r)g(r^{2})U(\vec{n})=R(r)U(\vec{n}),
\end{equation}
where $r=|\vec{A}|$, $\vec{n}$ is a unit vector on a $k$-dimensional
sphere, and $U=1/S_{k-1}(1)$ is a uniform distribution of vector
directions (or, in other words, a uniform distribution on the surface
of a $k$-dimensional ball). The distribution of directions can be
sampled by sampling a vector of $k$ normally distributed random numbers
and normalizing it to 1~\cite{Marsaglia1972,Muller1959}. In order
to sample the distribution of lengths, we have to do another change
of variable: $r^{2}\rightarrow x$, so that:
\begin{eqnarray}
R(x) & = & \frac{1}{2\sqrt{x}}S_{k-1}(\sqrt{x})g(x)\nonumber \\
 & = & \frac{1}{2\sqrt{x}}\frac{2\pi^{k/2}x^{(k-1)/2}}{\Gamma(k/2)}\frac{x^{N}}{\pi^{2}(N+1)!}\exp\left(-x\right)\nonumber \\
 & = & \frac{x^{N+1}}{\Gamma(N+2)}\exp\left(-x\right).
\end{eqnarray}
The result is exactly the gamma distribution with a shape parameter
$N+2$.

\subsection{Probabilistic violation of a Bell inequality}

\begin{figure}
\includegraphics{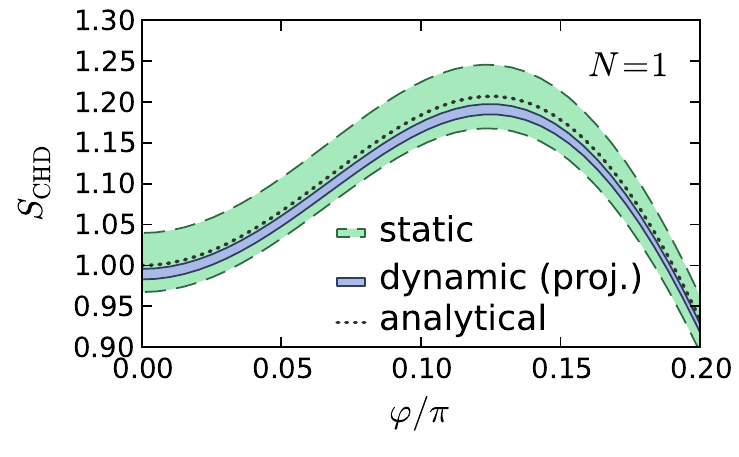}

\protect\caption{(Color online) Simulated moment-based CHD Bell violation $S_{\mathrm{CHD}}\left(\varphi\right)$
as a function of the relative polarizer angle $\varphi$ for one photon
pair using the positive P-distribution. Green dashed lines show the
result of static sampling with $2^{18}$ samples. Blue solid lines
show the results of the dynamic simulation with $2^{18}$ trajectories
at dimensionless time $\tau=0.1$. For each of the sampled states,
the filled region represents the range of the estimated error around
the mean of $S_{\mathrm{CHD}}(\varphi)$. The exact quantum mechanical
prediction of this value is represented by the grey dotted line.\label{fig:Bell-violation}}
\end{figure}

\begin{figure}
\includegraphics{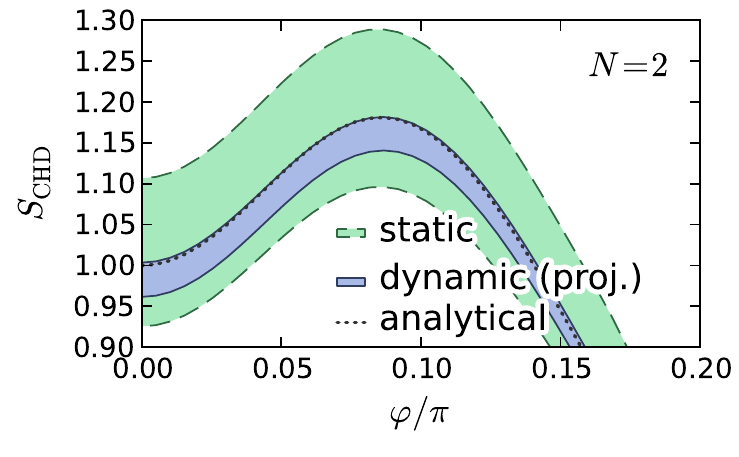}

\protect\caption{(Color online) Simulated moment-based Bell violation $S_{\mathrm{CHD}}\left(\varphi\right)$
as a function of the relative polarizer angle $\varphi$ for two photon
pairs using the positive P-distribution. Green dashed lines show the
result of static sampling with $2^{24}$ samples. Blue solid lines
show the results of a dynamic simulation with $2^{18}$ trajectories
at dimensionless time $\tau=0.1$. For each of the sampled states,
the filled region represents the range of the estimated error around
the mean of $S_{\mathrm{CHD}}(\varphi)$. The exact quantum mechanical
prediction of this value is represented by the grey dotted line.\label{fig:Bell-violation-2}}
\end{figure}

In Figs.~\ref{fig:Bell-violation} and~\ref{fig:Bell-violation-2}
we give computational results that show the probabilistic violation
of the bipartite $N$-photon CHD Bell inequality of Eqn.~(\ref{eq:BI_Cooperative}),
for polarized photons of the state~(\ref{eq:cooperative-state-1}).
Here we use $N=1$ and $N=2$ photons pairs respectively. In both
figures the dotted line corresponds to the quantum mechanical prediction.
Also plotted is the corresponding dynamical calculation, which is
explained in the next section. 

We have plotted the violation of $S_{CHD}$ defined in~(\ref{eq:BI_Cooperative}),
where the correlations $G^{IJ}$ were calculated using averages of
the corresponding phase space variable moments. In order to evaluate
the $N$-photon CHD Bell inequality, the relevant correlations are
given by:
\begin{eqnarray}
G^{IJ}(\theta,\phi,N) & = & \left\langle |\gamma_{+}|^{2I}|\delta_{+}|^{2J}\right\rangle _{P}\label{eq:G-in-phase-vars}\\
G^{IJ}(\theta,\infty,N) & = & \left\langle |\gamma_{+}|^{2I}\left(|\beta_{+}|^{2}+|\beta_{-}|^{2}\right)\right\rangle _{P}.\nonumber 
\end{eqnarray}

These results indicate a clear violation of the Bell inequality in
both the standard two-particle case ($N=1$) and the four-particle
case ($N=2$). This has also been observed experimentally~\cite{Howell2002}.
The computational results demonstrate a complete agreement with quantum
predictions up to the sampling error. This shows that these Bell violations
can certainly be simulated probabilistically.

The simulated Bell violations for this inequality are shown in Figs.~\ref{fig:Bell-violation}
for the $N=1$ case, and in Fig.~\ref{fig:Bell-violation-2} for
the $N=2$ case. This demonstrates a clear violation of a Bell inequality
using a probabilistic simulation, in both cases. The graphs include
results from a static simulation just of the Bell state, and also
from a dynamical simulation of a typical experiment using parametric
down-conversion, which will be explained next.\textcolor{red}{{} }We
used $2^{18}$ trajectories for dynamic simulations, $2^{18}$ samples
for the static sampling with $N=1$, and $2^{24}$ samples for the
static sampling with $N=2$ (to accommodate for the quickly growing
sampling error in the static case). The sampling error could be reduced
if we used more samples. 

We could instead have investigated the violations of other Bell inequalities
that use state probabilities rather than moments. All these inequalities
are operationally equivalent for the states used here, as explained
in Section II.C. However, state-projection calculations give larger
sampling errors than moments when using the standard `canonical' positive-P
distribution of Eqn.~(\ref{eq:cooperative-P}). This non-unique form
is easily computed, but it is more suitable for calculating moments
rather than probabilities. 

There are other expressions for the positive-P distribution, as well
as alternative representations like the general Gaussian representations~\cite{Corney2003}
which are better for sampling probabilities~\cite{Rosales-Zarate2011},
but are outside the scope of this article. The CH and CHSH state projection
Bell inequalities will be treated in the next section, which deals
with quantum dynamics.

\section{Dynamical Simulations\label{sec:Dynamical-Simulations}}

In this section we explain the model used to perform dynamical simulations
for the $N$-photon CHD-Bell inequality, together with the CH and
CHSH inequalities. In this latter case we also include the post-selection/
heralding process. This dynamical approach, as well as being more
physically realistic, has lower sampling errors than the static calculations.
The reason for this is that the dynamical equations generate a more
compact phase-space distribution, which is readily calculated using
stochastic methods. This improved sampling efficiency more than compensates
for the need to calculate time evolution, which is rather straightforward.

\subsection{\label{sub:Dynamical-Sim_NormCHSH}Dynamical simulations for the
$N$-photon CHD inequality}

In order to illustrate quantum dynamical simulations of violations
of Bell type inequalities using the positive P-representation, we
will consider the process of parametric down conversion described
earlier, which is modeled by the effective Hamiltonian of Eqn.~(\ref{eq:HamPDC-1}).
The positive P-representation provides a mapping that transforms the
time evolution of a density matrix into a set of phase space stochastic
equations. 

For the Hamiltonian of Eqn.~(\ref{eq:HamPDC-1}) we obtain the following
set of stochastic equations~\cite{Reid1986-violations,Dechoum2004}:
\begin{eqnarray}
d\alpha_{1} & = & \kappa E\beta_{1}^{+}dt+\sqrt{\kappa E}dW_{1}\nonumber \\
d\beta_{1} & = & \kappa E\alpha_{1}^{+}dt+\sqrt{\kappa E}dW_{1}^{*}\nonumber \\
d\alpha_{2} & = & \kappa E\beta_{2}^{+}dt+\sqrt{\kappa E}dW_{2}\nonumber \\
d\beta_{2} & = & \kappa E\alpha_{2}^{+}dt+\sqrt{\kappa E}dW_{2}^{*}\nonumber \\
d\alpha_{1}^{+} & = & \kappa E\beta_{1}dt+\sqrt{\kappa E}dW_{1}^{+}\nonumber \\
d\beta_{1}^{+} & = & \kappa E\alpha_{1}dt+\sqrt{\kappa E}\left(dW_{1}^{+}\right)^{*}\nonumber \\
d\alpha_{2^{+}} & = & \kappa E\beta_{2}dt+\sqrt{\kappa E}dW_{2}^{+}\nonumber \\
d\beta_{2}^{+} & = & \kappa E\alpha_{2}dt+\sqrt{\kappa E}\left(dW_{2}^{+}\right)^{*},\label{eq:SDE_PDC}
\end{eqnarray}
where the only non-vanishing correlations are 
\begin{equation}
\left\langle dW_{i}dW_{j}^{*}\right\rangle =\left\langle dW_{i}^{+}\left(dW_{j}^{+}\right)^{*}\right\rangle =dt\delta_{ij}.
\end{equation}

This set of Stratonovich stochastic equations of~(\ref{eq:SDE_PDC})
can be solved numerically in order to find the complex variables $\alpha_{i}\left(t\right)$,
$\beta_{i}\left(t\right)$ as a function of time. The rotated complex
phase-space variables $\gamma_{+}\left(t\right)$ and $\delta_{+}\left(t\right)$
are defined in Eqns.~(\ref{eq:cpm-gamma}) and~(\ref{eq:cpm-delta})
respectively. We use these complex variables to evaluate the intensity
correlations for one photon pair, described in Eqn.~(\ref{eq:Correlations}),
as well as the time evolution of the Bell type inequality of Eqn.~(\ref{eq:BI_Cooperative}). 

The expressions for the intensity correlations that we use are given,
as in the static case, by Eqn.~(\ref{eq:G-in-phase-vars}). We recall
that the dimensionless parametric interaction squeeze parameter $r$
is proportional to $\kappa Et$, and accordingly, we graph our results
against a dimensionless time $\tau=\kappa Et$.

In Figs.~\ref{fig:Bell-violation} and~\ref{fig:Bell-violation-2}
we show the results of dynamical simulations for $N=1$ and $N=2$
photons pairs respectively using the moment-based CHD Bell inequality.
These figures are plotted at $\tau\le0.25$, which we found was a
suitable time that minimizes the production of unwanted higher spin
multiple pairs.

Here we find that the sampling error in the dynamical case is smaller
that the static case. This is because the static distribution we used
has a simple existence theorem, but is non-unique, and usually does
not give the minimum variance possible. For all dynamical simulations
we used the central difference numerical algorithm~\cite{Drummond1991},
with coupling $\kappa E=1$ and time step $dt=2\times10^{-4}$, which
is sufficient to make discretization error negligible.

At very short times we observe a large sampling error. There is a
very clear physical reason for this. For these times the distribution
is dominated by the photonic vacuum state, giving a large sampling
error due to the fluctuations in the projection operator for the Bell
states. At times larger than about $\tau=0.1$, the onset of multiple
pair production occurs, which starts to reduce the Bell violation,
as we no longer have an ideal state.

\subsection{\label{sub:Dynamical-Sim_PDC}Dynamical simulations for the PDC process}

In order to obtain the time evolution of the violation of the Clauser-Horne
and CHSH Bell type inequalities using the positive P-representation,
we will now derive the appropriate operator mappings. As described
above, we use the positive P-representation in order to obtain the
complex variables $\alpha_{i}\left(t\right)$, $\beta_{i}\left(t\right)$
as a function of time and also the complex variables $\gamma_{i}\left(t\right)$,
$\delta_{i}\left(t\right)$, which are defined through the equations~(\ref{eq:cpm-gamma})
and~(\ref{eq:cpm-delta}) respectively.

\subsubsection{CH inequality}

The next step is the evaluation of each of the probabilities of the
CH inequality, Eqn.~(\ref{eq:CH}), as well as all the probabilities
of the CHSH inequality of Eqn.~(\ref{eq:chsh2}). Let us consider
one of these probabilities, for instance the probability of detecting
one photon in the up position at the polarizer with location A and
one photon in the ``up'' position at location B, $P_{++}^{AB}\left(\theta,\phi\right)$,
which is evaluated as follows:
\begin{eqnarray}
P_{++}^{AB}\left(\theta,\phi\right) & = & {\rm Tr}\left(\hat{\rho}\left|1100\right\rangle \left\langle 1100\right|\right)\nonumber \\
 & = & \int P\left(\vec{\gamma},\vec{\gamma}^{+}\right)\frac{\left\langle 1100\vert\vec{\gamma}\right\rangle \left\langle \left(\vec{\gamma}^{+}\right)^{*}\vert1100\right\rangle }{\left\langle \left(\vec{\gamma}^{+}\right)^{*}\vert\vec{\gamma}\right\rangle }\nonumber \\
 & = & \int P\left(\vec{\gamma},\vec{\gamma}^{+}\right)e^{-\vec{\gamma}^{+}\cdot\vec{\gamma}}\gamma_{1}\gamma_{1}^{+}\delta_{1}\delta_{1}^{+}d\vec{\gamma}d\vec{\gamma}^{+}.\label{eq:Probability-P}
\end{eqnarray}

Here $\vec{\gamma}=\left(\gamma_{1},\gamma_{2},\delta_{1},\delta_{2}\right)$
and $\vec{\gamma}^{+}=\left(\gamma_{1}^{+},\gamma_{2}^{+},\delta_{1}^{+},\delta_{2}^{+}\right)$.
The other probabilities are evaluated similarly. The marginal probabilities
$P_{+}^{A}\left(\theta\right)$ and $P_{+}^{B}\left(\phi\right)$
are evaluated as follows:
\begin{eqnarray}
P_{+}^{A}\left(\theta\right) & = & \int P\left(\vec{\gamma},\vec{\gamma}^{+}\right)e^{-\left(\gamma_{1}^{+}\gamma_{1}+\gamma_{2}^{+}\gamma_{2}\right)}\gamma_{1}^{+}\gamma_{1}d\vec{\gamma}d\vec{\gamma}^{+},\nonumber \\
P_{+}^{B}\left(\phi\right) & = & \int P\left(\vec{\gamma},\vec{\gamma}^{+}\right)e^{-\left(\delta_{1}^{+}\delta_{1}+\delta_{2}^{+}\delta_{2}\right)}\delta_{1}^{+}\delta_{1}d\vec{\gamma}d\vec{\gamma}^{+}.
\end{eqnarray}

To test the Clauser-Horne inequality, we evaluate the predictions
for $P_{++}^{AB}(\theta,\phi)$ and $P_{+}^{A}(\theta)$, for the
state created by the Hamiltonian~(\ref{eq:HamPDC-1}). Here, we have
defined the ``$+$'' event to be the detection of a single photon
at the up position, which means that the number of photons at the
down position does not need to be detected. This is the original formulation
of the CH inequality.

\begin{figure}
\includegraphics{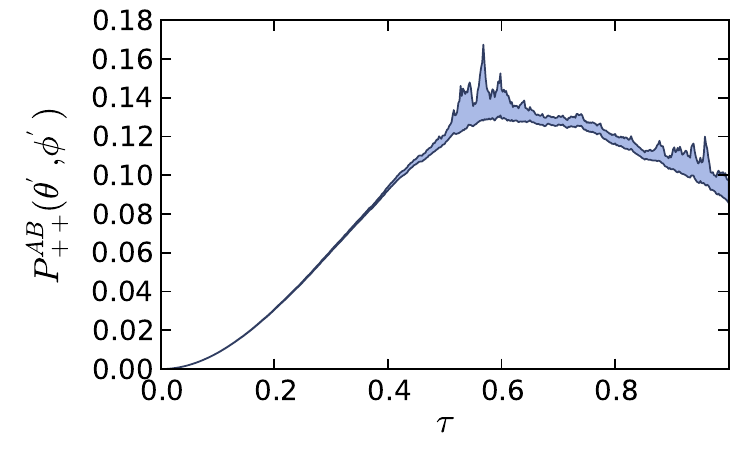}

\protect\caption{(Color online) Evolution of a single probability $P_{++}^{AB}(\theta^{\prime},\phi^{\prime})$
demonstrating that the sampling error increases after $t=0.5$.\label{fig:Probability}}
\end{figure}

In our simulations we noticed that after $\tau=0.5$ the sampling
error increased, as indicated by additional fluctuations after this
time in Fig.~\ref{fig:Probability}, which shows the evolution of
a single probability $P_{++}^{AB}(\theta^{\prime},\phi^{\prime})$.
The sampling error increase at long times is due to a larger proportion
of four and six particle states, and a correspondingly increased distribution
radius in phase-space. However, this is not optimal for Bell violations,
for which the short-time behavior is more physically important, as
experimentally these higher photon numbers are not utilized. Accordingly,
we take our samples at relatively short times with $\tau<0.25$ in
this paper, which is the most physically relevant time-scale. Experimentally
this corresponds to using a relatively short pump pulse or short interaction
distance, since this controls the evolution time.

For the purpose of calculation, we can choose our basis $a_{+}$ to
correspond to the mode axis $c_{+}$ defined by $\theta$ of polarizer
$A$. This amounts to putting $\theta=0$ in the calculation. We note
that the Hamiltonian $a_{+}^{\dagger}b_{+}^{\dagger}+a_{-}^{\dagger}b_{-}^{\dagger}$
is invariant under this type of rotation.

We can gain further insight into the solutions by writing the Hamiltonian
as $H=H_{1}+H_{2}$, where $H_{1\left(2\right)}=\kappa E\hat{a}_{1\left(2\right)}\hat{b}_{1\left(2\right)}+Hc$.
The evolution of the PDC is given by 
\begin{eqnarray}
|\psi\rangle & = & e^{-iHt/\hbar}|0\rangle\nonumber \\
 & = & e^{-iH_{1}t/\hbar}e^{-iH_{2}t/\hbar}|0\rangle\nonumber \\
 & = & (\sum_{n=0}c_{n}|n\rangle_{a_{1}}|n\rangle_{b_{1}})(\sum_{n=0}c_{n}|n\rangle_{a_{2}}|n\rangle_{b_{2}}),\label{eq:dyn}
\end{eqnarray}
where we have taken the initial state to be the multi-mode vacuum
state $|0\rangle$, which is the product of the vacuum states of each
of the four modes. Since $e^{-iH_{+}t/\hbar}|0\rangle$ is by definition
a two-mode squeezed state, we have used in the last line the well-known
result for the expansion of the two-mode squeezed state in terms of
the Fock number state basis. Here, the Fock state for mode $a$ is
denoted $|n\rangle_{a}$ and $c_{n}=\frac{x^{n}}{(1-x^{2})^{1/2}}$
where $x=\tanh r$~\cite{Hillery2006}. For small $r$, we can expand~(\ref{eq:dyn})
as explained in Section~\ref{sub:Bell-inequalities-PDC}: 
\begin{eqnarray}
|\psi\rangle & = & \tilde{c_{0}}|0\rangle+\tilde{c_{1}}|1\rangle+\ldots,\label{eq:exp}
\end{eqnarray}
where $|1\rangle$ is a Bell state and $\tilde{c}_{0}=c_{0}^{2}$,
$\tilde{c_{1}}=\sqrt{2}c_{0}c_{1}$. To gain an understanding of the
predictions for the Bell inequalities in this limit, we transform
to the modes of the measured basis $c_{+}$ and $d_{+}$: 
\begin{eqnarray}
|1\rangle & = & \frac{1}{\sqrt{2}}\left\{ (c_{+}^{\dagger}d_{+}^{\dagger}+c_{-}^{\dagger}d_{-}^{\dagger})\cos\varphi\right.\nonumber \\
 &  & \left.+(-c_{+}^{\dagger}d_{-}^{\dagger}+c_{-}^{\dagger}d_{+}^{\dagger})\sin\varphi\right\} |0\rangle,\label{eq:belltrans}
\end{eqnarray}
where $\varphi=\phi-\theta$. Then, we see that 
\begin{equation}
P_{++}^{AB}(\theta,\phi)=P_{--}^{AB}(\theta,\phi)=|\tilde{c}_{1}|^{2}\frac{1}{2}\cos^{2}(\theta-\phi),\label{eq:p++}
\end{equation}
 and $ $
\begin{equation}
P_{+-}^{AB}(\theta,\phi)=P_{-+}^{AB}(\theta,\phi)=|\tilde{c}_{1}|^{2}\frac{1}{2}\sin^{2}(\theta-\phi),\label{eq:P+-}
\end{equation}
 and the marginals are 
\begin{equation}
P_{+}^{A}(\theta)=P_{+}^{B}(\phi)=P_{-}^{A}(\theta)=P_{-}^{B}(\phi)=|\tilde{c}_{1}|^{2}/2.\label{eq:marg}
\end{equation}
We note that $|\tilde{c}_{1}|^{2}$ is the probability that the PDC
process generates the correlated photon-pair state, whereby a single
photon is incident on each detector. Choosing the usual case where
the angles $\theta$, $\phi$, $\theta'$ $\phi'$ increase sequentially
by $\pi/8$~\cite{Bell1964}, we see that for the CH inequality~(\ref{eq:CH})
the left side is $\left\{ |\tilde{c}_{1}|^{2}/2\right\} (3(\cos(\pi/4)/2+1/2)-(\cos3\pi/4+1)/2)=|\tilde{c}_{1}|^{2}(\sqrt{2}+1)/2$
whereas the right side is $ $$|\tilde{c}_{1}|^{2}$. Thus, in the
limit of small $r$, the ratio of the left to right side of the CH
inequality is predicted to be $(\sqrt{2}+1)/2$, which, being greater
than $1,$ violates the prediction of LHV theories. 

\begin{figure}
\includegraphics{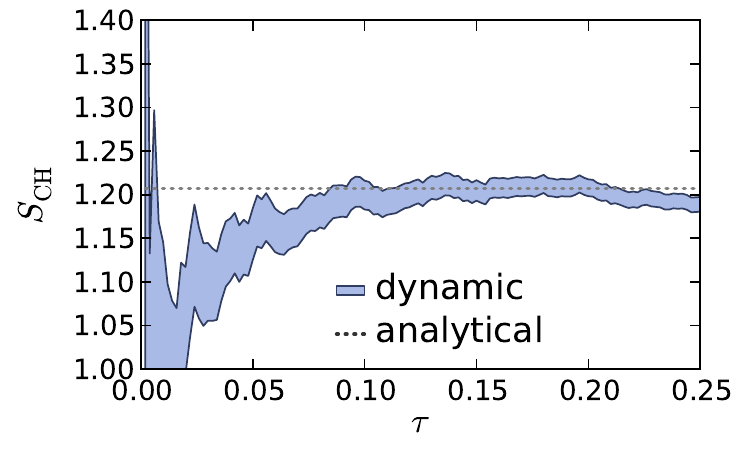}\protect\caption{(Color online) Evolution of the violation of the Clauser-Horne Bell
type inequality for the state generated by the parametric down conversion
process~(\ref{eq:HamPDC-1}): Plotted is the ratio $S_{\mathrm{CH}}$,
defined in Eqn.~(\ref{eq:Delta_CH}) for the angle choices $\theta=0$,
$\phi=\frac{\pi}{8}$, $\theta^{\prime}=\frac{\pi}{4}$ and $\phi^{\prime}=3\frac{\pi}{8}$.
Violation of the Bell inequality occurs when $S_{\mathrm{CH}}>1$.
The filled region represents the range of the estimated error around
the mean of $S_{\mathrm{CH}}$. The horizontal dotted line is the
expected value with no high-order components. Here we consider $2^{18}$
samples.\label{fig:Evolution_CH}}
\end{figure}

This result is indeed evident from our full solution, plotted in Fig.~\ref{fig:Evolution_CH},
where we have defined: 
\begin{eqnarray}
S_{\mathrm{CH}} & = & \frac{\tilde{P}(\theta,\phi)-\tilde{P}(\theta,\phi')+\tilde{P}(\theta',\phi)+\tilde{P}(\theta',\phi')}{P_{+}^{A}(\theta')+P_{+}^{B}(\phi)},\quad\label{eq:Delta_CH}
\end{eqnarray}
 where $\tilde{P}\equiv P_{++}^{AB}$. The violation of the CH inequality
as shown when $S_{\mathrm{CH}}>1$ is a rigorous (loophole-free) test
of LHV theories. 

In Fig.~\ref{fig:Evolution_CH} we show the time evolution simulations
of the violations of the Clauser-Horne inequality using $2^{18}$
samples. Since we are considering a ratio in this case, the results
are the same with or without the post-selection process. In fact post-selection
- which could introduce loop-holes in principle for other Bell inequalities
- simply has no effect on the measured data.

For higher $\tau$, correlated number states $|n\rangle|n\rangle$
where $n\geq2$ will also contribute to the statistics. Since we have
defined the ``$+$'' outcome to be that where a single photon is
detected at the ``up'' position, the violation of the CH inequality~(\ref{eq:CH})
diminishes. This is because a $+$ event at one detector can arise
from either the single or multi-photon Fock states. 

We note that the violation can be retained if one defines the $+$
outcome differently, to be detection of one photon in the ``up''
position and one photon in the ``down'' position, as we explained
above in Section~(\ref{sub:Bell-inequalities-PDC})~\cite{Reid2002}.
This latter definition amounts to the ``event selection'' method
of projection of the Bell state, and would prove a loophole-free test,
for larger times. Nonetheless, the probability of the actual measured
``$+$'' event becomes smaller in that case, and here we calculate
the behavior of the original CH inequality~(\ref{eq:CH}) to show
the dynamical evolution of the statistics.

\begin{figure}
\includegraphics{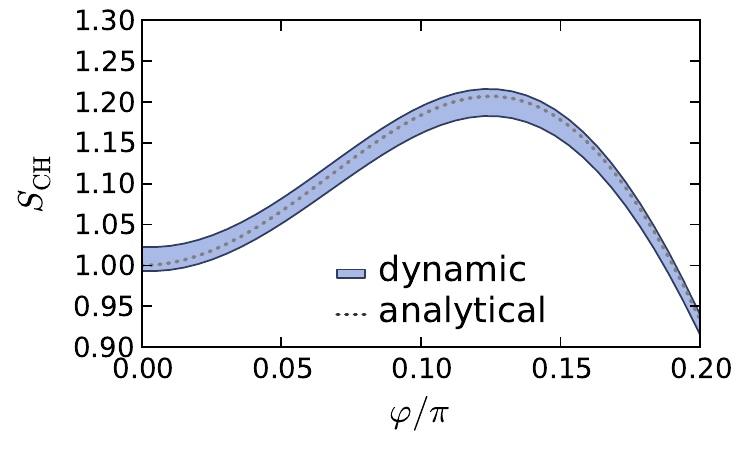}\protect\caption{(Color online) Angular dependence of the simulated Clauser-Horne Bell
type inequality for the state generated by the parametric down conversion
process~(\ref{eq:HamPDC-1}): Plotted is the ratio $S_{\mathrm{CH}}$,
defined in Eqn.~(\ref{eq:Delta_CH}) as a function of the relative
polarizer angle $\varphi=\phi-\theta=\phi^{\prime}-\theta^{\prime}=\theta^{\prime}-\phi$
at dimensionless time $\tau=0.1$. The filled region represents the
range of the estimated error around the mean of $S_{\mathrm{CH}}$.
Here we consider $2^{18}$ samples.\label{fig:CH_theta}}
\end{figure}

We find that these quantum dynamical simulations give a clear violation
of the Clauser-Horne inequality, which occurs when when $S_{\mathrm{CH}}>1$.
By choosing a particular time duration of $\tau=0.1$, we can examine
the detailed predictions for angular correlations with respect to
the relative polarizer angle $\varphi=\phi-\theta=\phi^{\prime}-\theta^{\prime}=\theta^{\prime}-\phi$.
This is shown in Fig.~\ref{fig:CH_theta}. 

\subsubsection{Sampling the CHSH inequality with and without post-selection}

In our simulations, we also consider how to simulate the experimental
post-selection/ heralding process, in which data is discarded in the
case where no photon is detected (the null-event). In order to do
this we consider a projection operator defined as:
\begin{equation}
\hat{P}=\hat{1}-\left|0\right\rangle \left\langle 0\right|.\label{eq:Proj}
\end{equation}
In this case the density matrix will be of the form: 
\begin{equation}
\hat{\rho}^{\prime}=\frac{\hat{P}^{\dagger}\hat{\rho}(t)\hat{P}}{Tr\left(\hat{P}^{\dagger}\hat{\rho}(t)\hat{P}\right)}.
\end{equation}
For the positive P-representation we will have a normalization given
by: 
\begin{eqnarray}
{\rm Tr}\left(\hat{P}^{\dagger}\frac{\left|\vec{\gamma}\right\rangle \langle\left(\vec{\gamma}^{+}\right)^{*}\vert}{\left\langle \left(\vec{\gamma}^{+}\right)^{*}\right|\left.\vec{\gamma}\right\rangle }\hat{P}\right) & = & \left\langle 1-e^{-\vec{\gamma}^{+}\cdot\vec{\gamma}}\right\rangle .\label{eq:Norm_Proj}
\end{eqnarray}
If we consider the post-selection process, the probabilities will
have normalization factor of the form of Eqn.~(\ref{eq:Norm_Proj}).
For instance, if we consider the post-selection process for the probability
$P_{++}^{AB}\left(\theta,\phi\right)$ that will be denoted as $P^{P}\left(\left|1100\right\rangle \right)$
we will obtain: 
\begin{equation}
P^{P}\left(\left|1100\right\rangle \right)=\frac{\left\langle \gamma_{1}\gamma_{1}^{+}\delta_{1}\delta_{1}^{+}e^{-\vec{\gamma}^{+}\cdot\vec{\gamma}}\right\rangle }{\left\langle 1-e^{-\vec{\gamma}^{+}\cdot\vec{\gamma}}\right\rangle }.
\end{equation}

\begin{figure}
\includegraphics{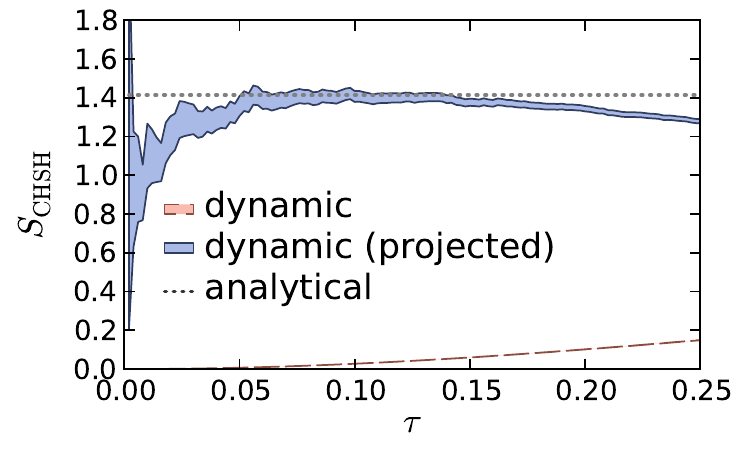}\protect\caption{(Color online) Evolution of the violation of the CHSH Bell inequality
for the state generated by the parametric down conversion process~(\ref{eq:HamPDC-1}):
Plotted is the ratio $S_{\mathrm{CHSH}}$, defined in Eqn.~(\ref{eq:DeltaCHSH})
for the angle choices $\theta=0$, $\phi=\frac{\pi}{8}$, $\theta^{\prime}=\frac{\pi}{4}$
and $\phi^{\prime}=3\frac{\pi}{8}$. Violation of the Bell inequality
occurs when $S_{\mathrm{CHSH}}>1$. The blue lines corresponds to
the case with post-selection/heralding where we consider the projector
operator defined in Eqn.~(\ref{eq:Proj}) and exclude the joint null
events from the statistics, while the red lines corresponds to the
simulation of the CHSH Bell-type inequality without post-selection.
In each case, the filled region represents the range of the sampled
error. The horizontal axis is the expected value at $t=0$.\label{fig:Evolution_CHSH}}
\end{figure}

Plotted in Fig.~\ref{fig:Evolution_CHSH}, are the predictions for
the CHSH inequality, both with and without post-selection. As previously,
we have defined 
\begin{eqnarray}
S_{\mathrm{CHSH}} & = & \frac{E(\theta,\phi)-E(\theta,\phi')+E(\theta',\phi)+E(\theta',\phi')}{2},\qquad\label{eq:DeltaCHSH}
\end{eqnarray}
 and we get violations when $S_{\mathrm{CHSH}}>1$. 

\begin{figure}
\includegraphics{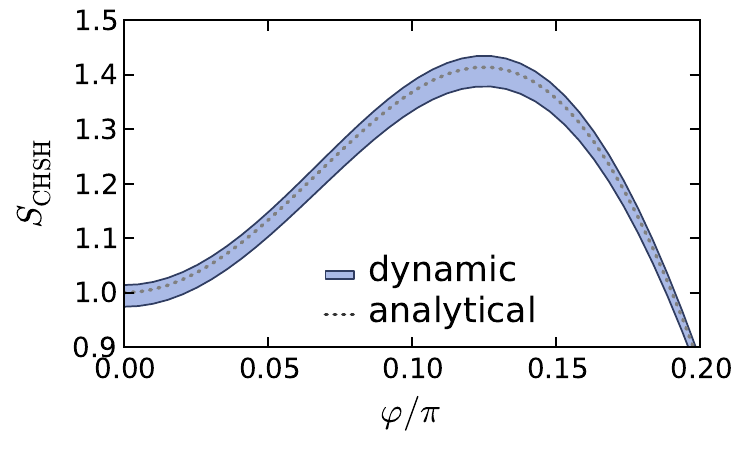}\protect\caption{(Color online) Angular dependence of the simulated CHSH Bell inequality
for the state generated by the parametric down conversion process~(\ref{eq:HamPDC-1}):
Plotted is the ratio $S_{\mathrm{CHSH}}$, defined in Eqn.~(\ref{eq:DeltaCHSH})
as a function of the relative polarizer angle $\varphi=\phi-\theta=\phi^{\prime}-\theta^{\prime}=\theta^{\prime}-\phi$
for dimensionless time $\tau=0.1$. In this case we consider the post-selection/
heralding process. The filled region represents the range of the sampled
error.\label{fig:CHSH_theta}}
\end{figure}

In Fig.~\ref{fig:CHSH_theta} we show the CHSH Bell type inequalities
as a function of the relative polarizer angle $\varphi=\phi-\theta=\phi^{\prime}-\theta^{\prime}=\theta^{\prime}-\phi$.
The simulations were performed at $\tau=0.1$. In the figure we show
the expected behavior for the CHSH inequality, including post-selection,
as a function of the angle. Also plotted in these figures in the corresponding
quantum mechanical prediction, showing excellent agreement.

\subsection{Extended simulations\label{sub:Extended-simulations}}

As an example of multi-mode problems which are of increasing interest
in physics, Bell violation PDC experiments in the laboratory have
much more complexity than the simple model we have used so far. Effects
not present in our model include multi-mode spatial propagation, nonlinearity
and loss. While it is the principle of probabilistic quantum simulation
of Bell violations that is of interest here, scalability is also important.
Therefore, we now show how our simulations can include such effects.

To treat these more realistic cases, we consider a waveguide based
gedanken experiment. As our model, we suppose that a single pump field
interacts with four down-converted waveguide modes, such that each
pair of polarization modes is generated in a single waveguide. This
is only one possible strategy to create a Bell violation. Laboratory
experiments use a variety of approaches, with different details in
each implementation.

While many methods are known experimentally, the extended model we
treat here is chosen as it is the closest to the four-mode treatment
given above, to allow a comparison. This waveguide proposal is actually
more complex than waveguides currently used. The stochastic equations
for this system have been obtained previously in simpler cases~\cite{Raymer1991,Werner1995},
and we extend this earlier analysis using the same techniques.

The main conclusion one reaches is that the ordinary stochastic differential
equations given in Eqn.~(\ref{eq:SDE_PDC}) are replaced by a very
similar set of partial stochastic differential equations for stochastic
fields. We assume for simplicity that all group velocities are equal
to $v$. The equations are~\cite{Raymer1991,Werner1995}:
\begin{eqnarray}
\left[\frac{\partial}{\partial z}+\frac{ik^{\prime\prime}}{2}\frac{\partial^{2}}{\partial t_{v}^{2}}\right]\Phi_{i}^{a} & = & -\gamma\Phi_{i}^{a}+\kappa^{*}\Psi\Phi_{i}^{b+}+\sqrt{\kappa^{*}\Psi}\zeta_{i}\nonumber \\
\left[\frac{\partial}{\partial z}+\frac{ik^{\prime\prime}}{2}\frac{\partial^{2}}{\partial t_{v}^{2}}\right]\Phi_{i}^{b} & = & -\gamma\Phi_{i}^{b}+\kappa^{*}\Psi\Phi_{i}^{a+}+\sqrt{\kappa^{*}\Psi}\zeta_{i}^{*}\nonumber \\
\left[\frac{\partial}{\partial z}+\frac{ik_{p}^{\prime\prime}}{2}\frac{\partial^{2}}{\partial t_{v}^{2}}\right]\Psi & = & -\gamma_{p}\Psi-\kappa\sum_{i}\Phi_{i}^{a}\Phi_{i}^{b}.
\end{eqnarray}

Here $\Psi$ is the stochastic pump field, while $\Phi_{i}^{a}$ and
$\Phi_{i}^{b}$ are the down-converted fields for $i=1,2$. There
are five equations for these fields, and five more independent equations
for the corresponding fields $\Psi^{+}$, $\Phi_{i}^{a+}$ and $\Phi_{i}^{b+}$.
All fields are flux amplitudes defined so that $\left\langle \Psi\Psi^{+}\right\rangle $
is the photon flux, with field units of $s^{-1/2}$. This normalization
is the most useful for the treatment of photon propagation.

The coordinate $z$ is the distance along the waveguide, $t_{v}=t-z/v$
is a moving frame time coordinate, $k^{\prime\prime}=d^{2}k/d\omega^{2}$
gives the group velocity dispersion, while $\gamma,\gamma_{p}$ are
the amplitude loss rates. The noise terms $\zeta_{i}$ are defined
as previously, except that they are now delta-correlated both in time
and space, rather than just in time. These equations include nonlinearity,
multi-mode dispersion and coupling to a dissipative reservoir describing
losses.

Apart from these modifications, solving these equations is very similar
to the original stochastic differential equations, with robust numerical
algorithms being available~\cite{Werner1997}. Waveguide experiments
of this type are known to be an efficient method of generating correlated
photons~\cite{U'Ren2004}, providing a useful alternative to atomic
cascade~\cite{Aspect1982-EPRB} or bulk crystal PDC~\cite{Fedrizzi2007}
experiments. 

We note that the mode indices $a,b$ describe mode polarizations,
while $i$ is a spatial mode index. The polarizations need to be swapped
with a polarizing beam-splitter in order to obtain the correlated
Bell state outputs that are required. Here, beam-splitting is a unitary
operation which is obtained through linear combinations of stochastic
terms. These issues are described in the original theoretical proposals
for PDC methods~\cite{Reid1986-violations,Shih1988}.

\section{\label{sec:Conclusions}Conclusions}

In his work on quantum computers~\cite{Feynman1982}, Feynman treated
an example of Bell states with correlations equivalent to the Bell
violating measurements we study. He showed that a probabilistic simulation
was not possible with simulations that sample the observed eigenvalues,
since they would be equivalent to hidden variable theories. This raises
the question of whether other types of probabilistic simulations can
be carried out for these states.

Our main result is very simple. There is no barrier to simulating
Bell violations probabilistically. The reason is that our phase-space
simulations do not use the operator eigenvalues. Instead, they employ
complex values whose averages and correlations are the same as the
known quantum correlations that violate a Bell inequality. For this
reason, such probabilistic quantum simulation methods are not hidden
variable theories, but are instead like quantum weak measurements~\cite{Aharonov1988}.
Therefore, they are not restricted to classical predictions that satisfy
Bell inequalities.

The simulations treated here were carried out in a number of ways,
either from known static phase-space distributions for the Bell state,
or dynamically. Time-dependent, dynamical simulations in fact are
simpler to implement than the static ones. We have investigated three
different types of Bell inequality, and have successfully simulated
them all, although the moment-based Bell inequalities are more well-suited
to the representation chosen here than ones using state projections
and quantum probabilities. The chief limitation of these methods is
their sampling error, which depends on the precise measurement simulated.
Another issue is the growth rate of sampling errors, which tend to
increase with time in a way that is not unlike the growth of error
in many classical dynamical systems.

Classical simulations of quantum systems commonly are thought to require
an exponentially large memory. With phase-space methods this requirement
disappears, since the phase-space dimension is simply proportional
to mode number. Instead, one must analyze the scaling of sampling
errors, which depends on the correlation order and the number of modes
and samples. This is analyzed in more detail in related investigations
into multipartite correlations~\cite{Opanchuk2014-bell-sim,Reid2014-60qubit},
which reached a size of $60$ qubits and $60$\textendash th order
moments. In these papers it was shown that exponentially large numbers
of samples can be required for simulations of mesoscopic quantum superpositions,
in cases where measured correlations have an order comparable to the
mode or qubit number. Such issues can result in exponentially long
simulation times. In practice, issues of inefficiency and noise limit
the order of correlations that can be measured in the laboratory~\cite{Rohde2014},
hence this is not usually a problem when simulating real experiments. 

If the measured correlations have a more realistic fixed order, as
in the present work, the scaling is much more favorable. Phase-space
methods have already been shown to provide accurate results even for
extremely large systems of bosons~\cite{Corney2006,Deuar2007}, with
such finite order moments. A more serious limitation of the present
method is the growth of sampling errors in time, which provides a
time horizon for accurate quantum predictions. We also emphasize that
the specific techniques used here are for bosonic, not fermionic systems.

In summary, the positive-P representation method was used because
it is a complete, positive representation of any bosonic quantum state,
with known procedures for obtaining dynamical quantum simulations.
This is especially useful in the case of the photonic parametric down-conversion
experiments widely utilized for Bell violations. The technique can
easily be expanded to treat more complex multi-mode Hamiltonians.
Other methods for phase-space mappings exist as well~\cite{Corney2003},
and some of these methods may converge even more rapidly. 
\begin{acknowledgments}
We wish to acknowledge funding from the Australian Research Council
Discovery Grants program.
\end{acknowledgments}
\bibliography{BellQsims}

\begin{thebibliography}{69}%
\makeatletter
\providecommand \@ifxundefined [1]{%
 \@ifx{#1\undefined}
}%
\providecommand \@ifnum [1]{%
 \ifnum #1\expandafter \@firstoftwo
 \else \expandafter \@secondoftwo
 \fi
}%
\providecommand \@ifx [1]{%
 \ifx #1\expandafter \@firstoftwo
 \else \expandafter \@secondoftwo
 \fi
}%
\providecommand \natexlab [1]{#1}%
\providecommand \enquote  [1]{``#1''}%
\providecommand \bibnamefont  [1]{#1}%
\providecommand \bibfnamefont [1]{#1}%
\providecommand \citenamefont [1]{#1}%
\providecommand \href@noop [0]{\@secondoftwo}%
\providecommand \href [0]{\begingroup \@sanitize@url \@href}%
\providecommand \@href[1]{\@@startlink{#1}\@@href}%
\providecommand \@@href[1]{\endgroup#1\@@endlink}%
\providecommand \@sanitize@url [0]{\catcode `\\12\catcode `\$12\catcode
  `\&12\catcode `\#12\catcode `\^12\catcode `\_12\catcode `\%12\relax}%
\providecommand \@@startlink[1]{}%
\providecommand \@@endlink[0]{}%
\providecommand \url  [0]{\begingroup\@sanitize@url \@url }%
\providecommand \@url [1]{\endgroup\@href {#1}{\urlprefix }}%
\providecommand \urlprefix  [0]{URL }%
\providecommand \Eprint [0]{\href }%
\providecommand \doibase [0]{http://dx.doi.org/}%
\providecommand \selectlanguage [0]{\@gobble}%
\providecommand \bibinfo  [0]{\@secondoftwo}%
\providecommand \bibfield  [0]{\@secondoftwo}%
\providecommand \translation [1]{[#1]}%
\providecommand \BibitemOpen [0]{}%
\providecommand \bibitemStop [0]{}%
\providecommand \bibitemNoStop [0]{.\EOS\space}%
\providecommand \EOS [0]{\spacefactor3000\relax}%
\providecommand \BibitemShut  [1]{\csname bibitem#1\endcsname}%
\let\auto@bib@innerbib\@empty
\bibitem [{\citenamefont {Dirac}(1929)}]{Dirac1929}%
  \BibitemOpen
  \bibfield  {author} {\bibinfo {author} {\bibfnamefont {P.~A.~M.}\
  \bibnamefont {Dirac}},\ }\href {\doibase 10.1098/rspa.1929.0094} {\bibfield
  {journal} {\bibinfo  {journal} {P. Roy. Soc. A.}\ }\textbf {\bibinfo {volume}
  {123}},\ \bibinfo {pages} {714} (\bibinfo {year} {1929})}\BibitemShut
  {NoStop}%
\bibitem [{\citenamefont {Feynman}(1982)}]{Feynman1982}%
  \BibitemOpen
  \bibfield  {author} {\bibinfo {author} {\bibfnamefont {R.~P.}\ \bibnamefont
  {Feynman}},\ }\href {\doibase 10.1007/BF02650179} {\bibfield  {journal}
  {\bibinfo  {journal} {Int. J. Theor. Phys.}\ }\textbf {\bibinfo {volume}
  {21}},\ \bibinfo {pages} {467} (\bibinfo {year} {1982})}\BibitemShut
  {NoStop}%
\bibitem [{\citenamefont {Haake}\ \emph {et~al.}(1979)\citenamefont {Haake},
  \citenamefont {King}, \citenamefont {Schr\"{o}der}, \citenamefont {Haus},\
  and\ \citenamefont {Glauber}}]{Haake1979}%
  \BibitemOpen
  \bibfield  {author} {\bibinfo {author} {\bibfnamefont {F.}~\bibnamefont
  {Haake}}, \bibinfo {author} {\bibfnamefont {H.}~\bibnamefont {King}},
  \bibinfo {author} {\bibfnamefont {G.}~\bibnamefont {Schr\"{o}der}}, \bibinfo
  {author} {\bibfnamefont {J.}~\bibnamefont {Haus}}, \ and\ \bibinfo {author}
  {\bibfnamefont {R.~J.}\ \bibnamefont {Glauber}},\ }\href {\doibase
  10.1103/PhysRevA.20.2047} {\bibfield  {journal} {\bibinfo  {journal} {Phys.
  Rev. A}\ }\textbf {\bibinfo {volume} {20}},\ \bibinfo {pages} {2047}
  (\bibinfo {year} {1979})}\BibitemShut {NoStop}%
\bibitem [{\citenamefont {Corney}\ \emph {et~al.}(2006)\citenamefont {Corney}
  \emph {et~al.}}]{Corney2006}%
  \BibitemOpen
  \bibfield  {author} {\bibinfo {author} {\bibfnamefont {J.~F.}\ \bibnamefont
  {Corney}} \emph {et~al.},\ }\href {\doibase 10.1103/PhysRevLett.97.023606}
  {\bibfield  {journal} {\bibinfo  {journal} {Phys. Rev. Lett.}\ }\textbf
  {\bibinfo {volume} {97}},\ \bibinfo {pages} {023606} (\bibinfo {year}
  {2006})}\BibitemShut {NoStop}%
\bibitem [{\citenamefont {Deuar}\ and\ \citenamefont
  {Drummond}(2007)}]{Deuar2007}%
  \BibitemOpen
  \bibfield  {author} {\bibinfo {author} {\bibfnamefont {P.~P.}\ \bibnamefont
  {Deuar}}\ and\ \bibinfo {author} {\bibfnamefont {P.~D.}\ \bibnamefont
  {Drummond}},\ }\href {\doibase 10.1103/PhysRevLett.98.120402} {\bibfield
  {journal} {\bibinfo  {journal} {Phys. Rev. Lett.}\ }\textbf {\bibinfo
  {volume} {98}},\ \bibinfo {pages} {120402} (\bibinfo {year}
  {2007})}\BibitemShut {NoStop}%
\bibitem [{\citenamefont {Alon}\ \emph {et~al.}(2008)\citenamefont {Alon},
  \citenamefont {Streltsov},\ and\ \citenamefont {Cederbaum}}]{Alon2008}%
  \BibitemOpen
  \bibfield  {author} {\bibinfo {author} {\bibfnamefont {O.~E.}\ \bibnamefont
  {Alon}}, \bibinfo {author} {\bibfnamefont {A.~I.}\ \bibnamefont {Streltsov}},
  \ and\ \bibinfo {author} {\bibfnamefont {L.~S.}\ \bibnamefont {Cederbaum}},\
  }\href {\doibase 10.1103/PhysRevA.77.033613} {\bibfield  {journal} {\bibinfo
  {journal} {Phys. Rev. A}\ }\textbf {\bibinfo {volume} {77}},\ \bibinfo
  {pages} {033613} (\bibinfo {year} {2008})}\BibitemShut {NoStop}%
\bibitem [{\citenamefont {Gambetta}\ \emph {et~al.}(2008)\citenamefont
  {Gambetta} \emph {et~al.}}]{Gambetta2008}%
  \BibitemOpen
  \bibfield  {author} {\bibinfo {author} {\bibfnamefont {J.}~\bibnamefont
  {Gambetta}} \emph {et~al.},\ }\href {\doibase 10.1103/PhysRevA.77.012112}
  {\bibfield  {journal} {\bibinfo  {journal} {Phys. Rev. A}\ }\textbf {\bibinfo
  {volume} {77}},\ \bibinfo {pages} {012112} (\bibinfo {year}
  {2008})}\BibitemShut {NoStop}%
\bibitem [{\citenamefont {Trotzky}\ \emph {et~al.}(2012)\citenamefont {Trotzky}
  \emph {et~al.}}]{Trotzky2012}%
  \BibitemOpen
  \bibfield  {author} {\bibinfo {author} {\bibfnamefont {S.}~\bibnamefont
  {Trotzky}} \emph {et~al.},\ }\href {\doibase 10.1038/nphys2232} {\bibfield
  {journal} {\bibinfo  {journal} {Nat. Phys.}\ }\textbf {\bibinfo {volume}
  {8}},\ \bibinfo {pages} {325} (\bibinfo {year} {2012})}\BibitemShut {NoStop}%
\bibitem [{\citenamefont {Cirac}\ and\ \citenamefont
  {Zoller}(2003)}]{Cirac2003}%
  \BibitemOpen
  \bibfield  {author} {\bibinfo {author} {\bibfnamefont {J.~I.}\ \bibnamefont
  {Cirac}}\ and\ \bibinfo {author} {\bibfnamefont {P.}~\bibnamefont {Zoller}},\
  }\href {\doibase 10.1126/science.1085130} {\bibfield  {journal} {\bibinfo
  {journal} {Science}\ }\textbf {\bibinfo {volume} {301}},\ \bibinfo {pages}
  {176} (\bibinfo {year} {2003})}\BibitemShut {NoStop}%
\bibitem [{\citenamefont {Jaksch}\ and\ \citenamefont
  {Zoller}(2005)}]{Jaksch2005}%
  \BibitemOpen
  \bibfield  {author} {\bibinfo {author} {\bibfnamefont {D.}~\bibnamefont
  {Jaksch}}\ and\ \bibinfo {author} {\bibfnamefont {P.}~\bibnamefont
  {Zoller}},\ }\href {\doibase 10.1016/j.aop.2004.09.010} {\bibfield  {journal}
  {\bibinfo  {journal} {Ann. Phys.}\ }\textbf {\bibinfo {volume} {315}},\
  \bibinfo {pages} {52} (\bibinfo {year} {2005})}\BibitemShut {NoStop}%
\bibitem [{\citenamefont {Buluta}\ and\ \citenamefont
  {Nori}(2009)}]{Buluta2009}%
  \BibitemOpen
  \bibfield  {author} {\bibinfo {author} {\bibfnamefont {I.}~\bibnamefont
  {Buluta}}\ and\ \bibinfo {author} {\bibfnamefont {F.}~\bibnamefont {Nori}},\
  }\href {\doibase 10.1126/science.1177838} {\bibfield  {journal} {\bibinfo
  {journal} {Science}\ }\textbf {\bibinfo {volume} {326}},\ \bibinfo {pages}
  {108} (\bibinfo {year} {2009})}\BibitemShut {NoStop}%
\bibitem [{\citenamefont {Islam}\ \emph {et~al.}(2011)\citenamefont {Islam}
  \emph {et~al.}}]{Islam2011}%
  \BibitemOpen
  \bibfield  {author} {\bibinfo {author} {\bibfnamefont {R.}~\bibnamefont
  {Islam}} \emph {et~al.},\ }\href {\doibase 10.1038/ncomms1374} {\bibfield
  {journal} {\bibinfo  {journal} {Nat. Commun.}\ }\textbf {\bibinfo {volume}
  {2}},\ \bibinfo {pages} {377} (\bibinfo {year} {2011})}\BibitemShut {NoStop}%
\bibitem [{\citenamefont {Georgescu}\ \emph {et~al.}(2014)\citenamefont
  {Georgescu}, \citenamefont {Ashhab},\ and\ \citenamefont
  {Nori}}]{Georgescu2014}%
  \BibitemOpen
  \bibfield  {author} {\bibinfo {author} {\bibfnamefont {I.~M.}\ \bibnamefont
  {Georgescu}}, \bibinfo {author} {\bibfnamefont {S.}~\bibnamefont {Ashhab}}, \
  and\ \bibinfo {author} {\bibfnamefont {F.}~\bibnamefont {Nori}},\ }\href
  {\doibase 10.1103/RevModPhys.86.153} {\bibfield  {journal} {\bibinfo
  {journal} {Rev. Mod. Phys.}\ }\textbf {\bibinfo {volume} {86}},\ \bibinfo
  {pages} {153} (\bibinfo {year} {2014})}\BibitemShut {NoStop}%
\bibitem [{\citenamefont {Lanyon}\ \emph {et~al.}(2011)\citenamefont {Lanyon}
  \emph {et~al.}}]{Lanyon2011}%
  \BibitemOpen
  \bibfield  {author} {\bibinfo {author} {\bibfnamefont {B.~P.}\ \bibnamefont
  {Lanyon}} \emph {et~al.},\ }\href {\doibase 10.1126/science.1208001}
  {\bibfield  {journal} {\bibinfo  {journal} {Science}\ }\textbf {\bibinfo
  {volume} {334}},\ \bibinfo {pages} {57} (\bibinfo {year} {2011})}\BibitemShut
  {NoStop}%
\bibitem [{\citenamefont {Opanchuk}\ \emph {et~al.}(2013)\citenamefont
  {Opanchuk}, \citenamefont {Polkinghorne}, \citenamefont {Fialko},
  \citenamefont {Brand},\ and\ \citenamefont
  {Drummond}}]{Opanchuk2013-early-universe}%
  \BibitemOpen
  \bibfield  {author} {\bibinfo {author} {\bibfnamefont {B.}~\bibnamefont
  {Opanchuk}}, \bibinfo {author} {\bibfnamefont {R.}~\bibnamefont
  {Polkinghorne}}, \bibinfo {author} {\bibfnamefont {O.}~\bibnamefont
  {Fialko}}, \bibinfo {author} {\bibfnamefont {J.}~\bibnamefont {Brand}}, \
  and\ \bibinfo {author} {\bibfnamefont {P.~D.}\ \bibnamefont {Drummond}},\
  }\href {\doibase 10.1002/andp.201300113} {\bibfield  {journal} {\bibinfo
  {journal} {Ann. Phys.}\ }\textbf {\bibinfo {volume} {525}},\ \bibinfo {pages}
  {866} (\bibinfo {year} {2013})}\BibitemShut {NoStop}%
\bibitem [{\citenamefont {Drummond}\ \emph {et~al.}(1993)\citenamefont
  {Drummond}, \citenamefont {Shelby}, \citenamefont {Friberg},\ and\
  \citenamefont {Yamamoto}}]{Drummond1993-solitons}%
  \BibitemOpen
  \bibfield  {author} {\bibinfo {author} {\bibfnamefont {P.~D.}\ \bibnamefont
  {Drummond}}, \bibinfo {author} {\bibfnamefont {R.~M.}\ \bibnamefont
  {Shelby}}, \bibinfo {author} {\bibfnamefont {S.~R.}\ \bibnamefont {Friberg}},
  \ and\ \bibinfo {author} {\bibfnamefont {Y.}~\bibnamefont {Yamamoto}},\
  }\href {\doibase 10.1038/365307a0} {\bibfield  {journal} {\bibinfo  {journal}
  {Nature}\ }\textbf {\bibinfo {volume} {365}},\ \bibinfo {pages} {307}
  (\bibinfo {year} {1993})}\BibitemShut {NoStop}%
\bibitem [{\citenamefont {Dechoum}\ \emph {et~al.}(2004)\citenamefont
  {Dechoum}, \citenamefont {Drummond}, \citenamefont {Chaturvedi},\ and\
  \citenamefont {Reid}}]{Dechoum2004}%
  \BibitemOpen
  \bibfield  {author} {\bibinfo {author} {\bibfnamefont {K.}~\bibnamefont
  {Dechoum}}, \bibinfo {author} {\bibfnamefont {P.~D.}\ \bibnamefont
  {Drummond}}, \bibinfo {author} {\bibfnamefont {S.}~\bibnamefont
  {Chaturvedi}}, \ and\ \bibinfo {author} {\bibfnamefont {M.~D.}\ \bibnamefont
  {Reid}},\ }\href {\doibase 10.1103/PhysRevA.70.053807} {\bibfield  {journal}
  {\bibinfo  {journal} {Phys. Rev. A}\ }\textbf {\bibinfo {volume} {70}},\
  \bibinfo {pages} {53807} (\bibinfo {year} {2004})}\BibitemShut {NoStop}%
\bibitem [{\citenamefont {Deuar}\ \emph {et~al.}(2011)\citenamefont {Deuar},
  \citenamefont {Chwede\'nczuk}, \citenamefont {Trippenbach},\ and\
  \citenamefont {Zi\'n}}]{Deuar2011}%
  \BibitemOpen
  \bibfield  {author} {\bibinfo {author} {\bibfnamefont {P.~P.}\ \bibnamefont
  {Deuar}}, \bibinfo {author} {\bibfnamefont {J.}~\bibnamefont
  {Chwede\'nczuk}}, \bibinfo {author} {\bibfnamefont {M.}~\bibnamefont
  {Trippenbach}}, \ and\ \bibinfo {author} {\bibfnamefont {P.}~\bibnamefont
  {Zi\'n}},\ }\href {\doibase 10.1103/PhysRevA.83.063625} {\bibfield  {journal}
  {\bibinfo  {journal} {Phys. Rev. A}\ }\textbf {\bibinfo {volume} {83}},\
  \bibinfo {pages} {063625} (\bibinfo {year} {2011})}\BibitemShut {NoStop}%
\bibitem [{\citenamefont {Krachmalnicoff}\ \emph {et~al.}(2010)\citenamefont
  {Krachmalnicoff} \emph {et~al.}}]{Krachmalnicoff2010}%
  \BibitemOpen
  \bibfield  {author} {\bibinfo {author} {\bibfnamefont {V.}~\bibnamefont
  {Krachmalnicoff}} \emph {et~al.},\ }\href {\doibase
  10.1103/PhysRevLett.104.150402} {\bibfield  {journal} {\bibinfo  {journal}
  {Phys. Rev. Lett.}\ }\textbf {\bibinfo {volume} {104}},\ \bibinfo {pages}
  {150402} (\bibinfo {year} {2010})}\BibitemShut {NoStop}%
\bibitem [{\citenamefont {Clauser}\ and\ \citenamefont
  {Horne}(1974)}]{Clauser1974}%
  \BibitemOpen
  \bibfield  {author} {\bibinfo {author} {\bibfnamefont {J.~F.}\ \bibnamefont
  {Clauser}}\ and\ \bibinfo {author} {\bibfnamefont {M.~A.}\ \bibnamefont
  {Horne}},\ }\href {\doibase 10.1103/PhysRevD.10.526} {\bibfield  {journal}
  {\bibinfo  {journal} {Phys. Rev. D}\ }\textbf {\bibinfo {volume} {10}},\
  \bibinfo {pages} {526} (\bibinfo {year} {1974})}\BibitemShut {NoStop}%
\bibitem [{\citenamefont {Bell}(1964)}]{Bell1964}%
  \BibitemOpen
  \bibfield  {author} {\bibinfo {author} {\bibfnamefont {J.~S.}\ \bibnamefont
  {Bell}},\ }\href
  {http://philoscience.unibe.ch/documents/TexteHS10/bell1964epr.pdf} {\bibfield
   {journal} {\bibinfo  {journal} {Physics}\ }\textbf {\bibinfo {volume} {1}},\
  \bibinfo {pages} {195} (\bibinfo {year} {1964})}\BibitemShut {NoStop}%
\bibitem [{\citenamefont {Clauser}\ \emph {et~al.}(1969)\citenamefont
  {Clauser}, \citenamefont {Horne}, \citenamefont {Shimony},\ and\
  \citenamefont {Holt}}]{Clauser1969}%
  \BibitemOpen
  \bibfield  {author} {\bibinfo {author} {\bibfnamefont {J.~F.}\ \bibnamefont
  {Clauser}}, \bibinfo {author} {\bibfnamefont {M.~A.}\ \bibnamefont {Horne}},
  \bibinfo {author} {\bibfnamefont {A.}~\bibnamefont {Shimony}}, \ and\
  \bibinfo {author} {\bibfnamefont {R.~A.}\ \bibnamefont {Holt}},\ }\href
  {\doibase 10.1103/PhysRevLett.23.880} {\bibfield  {journal} {\bibinfo
  {journal} {Phys. Rev. Lett.}\ }\textbf {\bibinfo {volume} {23}},\ \bibinfo
  {pages} {880} (\bibinfo {year} {1969})}\BibitemShut {NoStop}%
\bibitem [{\citenamefont {Drummond}(1983)}]{Drummond1983}%
  \BibitemOpen
  \bibfield  {author} {\bibinfo {author} {\bibfnamefont {P.~D.}\ \bibnamefont
  {Drummond}},\ }\href {\doibase 10.1103/PhysRevLett.50.1407} {\bibfield
  {journal} {\bibinfo  {journal} {Phys. Rev. Lett.}\ }\textbf {\bibinfo
  {volume} {50}},\ \bibinfo {pages} {1407} (\bibinfo {year}
  {1983})}\BibitemShut {NoStop}%
\bibitem [{\citenamefont {Reid}\ \emph {et~al.}(2002)\citenamefont {Reid},
  \citenamefont {Munro},\ and\ \citenamefont {{De Martini}}}]{Reid2002}%
  \BibitemOpen
  \bibfield  {author} {\bibinfo {author} {\bibfnamefont {M.~D.}\ \bibnamefont
  {Reid}}, \bibinfo {author} {\bibfnamefont {W.~J.}\ \bibnamefont {Munro}}, \
  and\ \bibinfo {author} {\bibfnamefont {F.}~\bibnamefont {{De Martini}}},\
  }\href {\doibase 10.1103/PhysRevA.66.033801} {\bibfield  {journal} {\bibinfo
  {journal} {Phys. Rev. A}\ }\textbf {\bibinfo {volume} {66}},\ \bibinfo
  {pages} {033801} (\bibinfo {year} {2002})}\BibitemShut {NoStop}%
\bibitem [{\citenamefont {Drummond}\ \emph {et~al.}(2014)\citenamefont
  {Drummond}, \citenamefont {Opanchuk}, \citenamefont {Rosales-Z\'{a}rate},\
  and\ \citenamefont {Reid}}]{Drummond2014-bell-sim}%
  \BibitemOpen
  \bibfield  {author} {\bibinfo {author} {\bibfnamefont {P.~D.}\ \bibnamefont
  {Drummond}}, \bibinfo {author} {\bibfnamefont {B.}~\bibnamefont {Opanchuk}},
  \bibinfo {author} {\bibfnamefont {L.~E.~C.}\ \bibnamefont
  {Rosales-Z\'{a}rate}}, \ and\ \bibinfo {author} {\bibfnamefont {M.~D.}\
  \bibnamefont {Reid}},\ }\href {\doibase 10.1088/0031-8949/2014/T160/014009}
  {\bibfield  {journal} {\bibinfo  {journal} {Phys. Scripta}\ }\textbf
  {\bibinfo {volume} {T160}},\ \bibinfo {pages} {014009} (\bibinfo {year}
  {2014})}\BibitemShut {NoStop}%
\bibitem [{\citenamefont {Reid}\ and\ \citenamefont
  {Walls}(1986)}]{Reid1986-violations}%
  \BibitemOpen
  \bibfield  {author} {\bibinfo {author} {\bibfnamefont {M.~D.}\ \bibnamefont
  {Reid}}\ and\ \bibinfo {author} {\bibfnamefont {D.~F.}\ \bibnamefont
  {Walls}},\ }\href {\doibase 10.1103/PhysRevA.34.1260} {\bibfield  {journal}
  {\bibinfo  {journal} {Phys. Rev. A}\ }\textbf {\bibinfo {volume} {34}},\
  \bibinfo {pages} {1260} (\bibinfo {year} {1986})}\BibitemShut {NoStop}%
\bibitem [{\citenamefont {Shih}\ and\ \citenamefont {Alley}(1988)}]{Shih1988}%
  \BibitemOpen
  \bibfield  {author} {\bibinfo {author} {\bibfnamefont {Y.~H.}\ \bibnamefont
  {Shih}}\ and\ \bibinfo {author} {\bibfnamefont {C.~O.}\ \bibnamefont
  {Alley}},\ }\href {\doibase 10.1103/PhysRevLett.61.2921} {\bibfield
  {journal} {\bibinfo  {journal} {Phys. Rev. Lett.}\ }\textbf {\bibinfo
  {volume} {61}},\ \bibinfo {pages} {2921} (\bibinfo {year}
  {1988})}\BibitemShut {NoStop}%
\bibitem [{\citenamefont {Ou}\ and\ \citenamefont {Mandel}(1988)}]{Ou1988}%
  \BibitemOpen
  \bibfield  {author} {\bibinfo {author} {\bibfnamefont {Z.~Y.}\ \bibnamefont
  {Ou}}\ and\ \bibinfo {author} {\bibfnamefont {L.}~\bibnamefont {Mandel}},\
  }\href {\doibase 10.1103/PhysRevLett.61.50} {\bibfield  {journal} {\bibinfo
  {journal} {Phys. Rev. Lett.}\ }\textbf {\bibinfo {volume} {61}},\ \bibinfo
  {pages} {50} (\bibinfo {year} {1988})}\BibitemShut {NoStop}%
\bibitem [{\citenamefont {Kwiat}\ \emph {et~al.}(1995)\citenamefont {Kwiat}
  \emph {et~al.}}]{Kwiat1995}%
  \BibitemOpen
  \bibfield  {author} {\bibinfo {author} {\bibfnamefont {P.~G.}\ \bibnamefont
  {Kwiat}} \emph {et~al.},\ }\href {\doibase 10.1103/PhysRevLett.75.4337}
  {\bibfield  {journal} {\bibinfo  {journal} {Phys. Rev. Lett.}\ }\textbf
  {\bibinfo {volume} {75}},\ \bibinfo {pages} {4337} (\bibinfo {year}
  {1995})}\BibitemShut {NoStop}%
\bibitem [{\citenamefont {Drummond}\ and\ \citenamefont
  {Gardiner}(1980)}]{Drummond1980}%
  \BibitemOpen
  \bibfield  {author} {\bibinfo {author} {\bibfnamefont {P.~D.}\ \bibnamefont
  {Drummond}}\ and\ \bibinfo {author} {\bibfnamefont {C.~W.}\ \bibnamefont
  {Gardiner}},\ }\href {\doibase 10.1088/0305-4470/13/7/018} {\bibfield
  {journal} {\bibinfo  {journal} {J. Phys. A: Math. Gen.}\ }\textbf {\bibinfo
  {volume} {13}},\ \bibinfo {pages} {2353} (\bibinfo {year}
  {1980})}\BibitemShut {NoStop}%
\bibitem [{\citenamefont {Hillery}\ \emph {et~al.}(1984)\citenamefont
  {Hillery}, \citenamefont {O'Connell}, \citenamefont {Scully},\ and\
  \citenamefont {Wigner}}]{Hillery1984}%
  \BibitemOpen
  \bibfield  {author} {\bibinfo {author} {\bibfnamefont {M.}~\bibnamefont
  {Hillery}}, \bibinfo {author} {\bibfnamefont {R.~F.}\ \bibnamefont
  {O'Connell}}, \bibinfo {author} {\bibfnamefont {M.~O.}\ \bibnamefont
  {Scully}}, \ and\ \bibinfo {author} {\bibfnamefont {E.~P.}\ \bibnamefont
  {Wigner}},\ }\href {\doibase 10.1016/0370-1573(84)90160-1} {\bibfield
  {journal} {\bibinfo  {journal} {Phys. Rep.}\ }\textbf {\bibinfo {volume}
  {106}},\ \bibinfo {pages} {121} (\bibinfo {year} {1984})}\BibitemShut
  {NoStop}%
\bibitem [{\citenamefont {Giustina}\ \emph {et~al.}(2013)\citenamefont
  {Giustina} \emph {et~al.}}]{Giustina2013}%
  \BibitemOpen
  \bibfield  {author} {\bibinfo {author} {\bibfnamefont {M.}~\bibnamefont
  {Giustina}} \emph {et~al.},\ }\href {\doibase 10.1038/nature12012} {\bibfield
   {journal} {\bibinfo  {journal} {Nature}\ }\textbf {\bibinfo {volume}
  {497}},\ \bibinfo {pages} {227} (\bibinfo {year} {2013})}\BibitemShut
  {NoStop}%
\bibitem [{\citenamefont {Christensen}\ \emph {et~al.}(2013)\citenamefont
  {Christensen} \emph {et~al.}}]{Christensen2013}%
  \BibitemOpen
  \bibfield  {author} {\bibinfo {author} {\bibfnamefont {B.~G.}\ \bibnamefont
  {Christensen}} \emph {et~al.},\ }\href {\doibase
  10.1103/PhysRevLett.111.130406} {\bibfield  {journal} {\bibinfo  {journal}
  {Phys. Rev. Lett.}\ }\textbf {\bibinfo {volume} {111}},\ \bibinfo {pages}
  {130406} (\bibinfo {year} {2013})}\BibitemShut {NoStop}%
\bibitem [{\citenamefont {Opanchuk}\ \emph {et~al.}(2014)\citenamefont
  {Opanchuk}, \citenamefont {Rosales-Z\'{a}rate}, \citenamefont {Reid},\ and\
  \citenamefont {Drummond}}]{Opanchuk2014-bell-sim}%
  \BibitemOpen
  \bibfield  {author} {\bibinfo {author} {\bibfnamefont {B.}~\bibnamefont
  {Opanchuk}}, \bibinfo {author} {\bibfnamefont {L.~E.~C.}\ \bibnamefont
  {Rosales-Z\'{a}rate}}, \bibinfo {author} {\bibfnamefont {M.~D.}\ \bibnamefont
  {Reid}}, \ and\ \bibinfo {author} {\bibfnamefont {P.~D.}\ \bibnamefont
  {Drummond}},\ }\href {\doibase 10.1016/j.physleta.2014.01.045} {\bibfield
  {journal} {\bibinfo  {journal} {Phys. Lett. A}\ }\textbf {\bibinfo {volume}
  {378}},\ \bibinfo {pages} {946} (\bibinfo {year} {2014})}\BibitemShut
  {NoStop}%
\bibitem [{\citenamefont {Aharonov}\ \emph {et~al.}(1988)\citenamefont
  {Aharonov}, \citenamefont {Albert},\ and\ \citenamefont
  {Vaidman}}]{Aharonov1988}%
  \BibitemOpen
  \bibfield  {author} {\bibinfo {author} {\bibfnamefont {Y.}~\bibnamefont
  {Aharonov}}, \bibinfo {author} {\bibfnamefont {D.~Z.}\ \bibnamefont
  {Albert}}, \ and\ \bibinfo {author} {\bibfnamefont {L.}~\bibnamefont
  {Vaidman}},\ }\href {\doibase 10.1103/PhysRevLett.60.1351} {\bibfield
  {journal} {\bibinfo  {journal} {Phys. Rev. Lett.}\ }\textbf {\bibinfo
  {volume} {60}},\ \bibinfo {pages} {1351} (\bibinfo {year}
  {1988})}\BibitemShut {NoStop}%
\bibitem [{\citenamefont {Clauser}\ and\ \citenamefont
  {Shimony}(1978)}]{Clauser1978}%
  \BibitemOpen
  \bibfield  {author} {\bibinfo {author} {\bibfnamefont {J.~F.}\ \bibnamefont
  {Clauser}}\ and\ \bibinfo {author} {\bibfnamefont {A.}~\bibnamefont
  {Shimony}},\ }\href {\doibase 10.1088/0034-4885/41/12/002} {\bibfield
  {journal} {\bibinfo  {journal} {Rep. Prog. Phys.}\ }\textbf {\bibinfo
  {volume} {41}},\ \bibinfo {pages} {1881} (\bibinfo {year}
  {1978})}\BibitemShut {NoStop}%
\bibitem [{\citenamefont {Aspect}\ \emph
  {et~al.}(1982{\natexlab{a}})\citenamefont {Aspect}, \citenamefont
  {Grangier},\ and\ \citenamefont {Roger}}]{Aspect1982-EPRB}%
  \BibitemOpen
  \bibfield  {author} {\bibinfo {author} {\bibfnamefont {A.}~\bibnamefont
  {Aspect}}, \bibinfo {author} {\bibfnamefont {P.}~\bibnamefont {Grangier}}, \
  and\ \bibinfo {author} {\bibfnamefont {G.}~\bibnamefont {Roger}},\ }\href
  {\doibase 10.1103/PhysRevLett.49.91} {\bibfield  {journal} {\bibinfo
  {journal} {Phys. Rev. Lett.}\ }\textbf {\bibinfo {volume} {49}},\ \bibinfo
  {pages} {91} (\bibinfo {year} {1982}{\natexlab{a}})}\BibitemShut {NoStop}%
\bibitem [{\citenamefont {D'Espagnat}(1971)}]{D'Espagnat1971}%
  \BibitemOpen
  \bibinfo {editor} {\bibfnamefont {B.}~\bibnamefont {D'Espagnat}},\ ed.,\
  \href@noop {} {\emph {\bibinfo {title} {{Foundations of Quantum
  Mechanics}}}}\ (\bibinfo  {publisher} {Academic Press},\ \bibinfo {address}
  {New York},\ \bibinfo {year} {1971})\BibitemShut {NoStop}%
\bibitem [{\citenamefont {Cirel'son}(1980)}]{Cirel'son1980}%
  \BibitemOpen
  \bibfield  {author} {\bibinfo {author} {\bibfnamefont {B.~S.}\ \bibnamefont
  {Cirel'son}},\ }\href {\doibase 10.1007/BF00417500} {\bibfield  {journal}
  {\bibinfo  {journal} {Lett. Math. Phys.}\ }\textbf {\bibinfo {volume} {4}},\
  \bibinfo {pages} {93} (\bibinfo {year} {1980})}\BibitemShut {NoStop}%
\bibitem [{\citenamefont {Aspect}\ \emph
  {et~al.}(1982{\natexlab{b}})\citenamefont {Aspect}, \citenamefont
  {Dalibard},\ and\ \citenamefont {Roger}}]{Aspect1982-time-varying}%
  \BibitemOpen
  \bibfield  {author} {\bibinfo {author} {\bibfnamefont {A.}~\bibnamefont
  {Aspect}}, \bibinfo {author} {\bibfnamefont {J.}~\bibnamefont {Dalibard}}, \
  and\ \bibinfo {author} {\bibfnamefont {G.}~\bibnamefont {Roger}},\ }\href
  {\doibase 10.1103/PhysRevLett.49.1804} {\bibfield  {journal} {\bibinfo
  {journal} {Phys. Rev. Lett.}\ }\textbf {\bibinfo {volume} {49}},\ \bibinfo
  {pages} {1804} (\bibinfo {year} {1982}{\natexlab{b}})}\BibitemShut {NoStop}%
\bibitem [{\citenamefont {Eberhard}(1993)}]{Eberhard1993}%
  \BibitemOpen
  \bibfield  {author} {\bibinfo {author} {\bibfnamefont {P.}~\bibnamefont
  {Eberhard}},\ }\href {\doibase 10.1103/PhysRevA.47.R747} {\bibfield
  {journal} {\bibinfo  {journal} {Phys. Rev. A}\ }\textbf {\bibinfo {volume}
  {47}},\ \bibinfo {pages} {R747} (\bibinfo {year} {1993})}\BibitemShut
  {NoStop}%
\bibitem [{\citenamefont {Cabello}(2007)}]{Cabello2007}%
  \BibitemOpen
  \bibfield  {author} {\bibinfo {author} {\bibfnamefont {A.}~\bibnamefont
  {Cabello}},\ }\href {\doibase 10.1103/PhysRevLett.98.220402} {\bibfield
  {journal} {\bibinfo  {journal} {Phys. Rev. Lett.}\ }\textbf {\bibinfo
  {volume} {98}},\ \bibinfo {pages} {220402} (\bibinfo {year}
  {2007})}\BibitemShut {NoStop}%
\bibitem [{\citenamefont {Brunner}\ \emph {et~al.}(2007)\citenamefont
  {Brunner}, \citenamefont {Gisin}, \citenamefont {Scarani},\ and\
  \citenamefont {Simon}}]{Brunner2007}%
  \BibitemOpen
  \bibfield  {author} {\bibinfo {author} {\bibfnamefont {N.}~\bibnamefont
  {Brunner}}, \bibinfo {author} {\bibfnamefont {N.}~\bibnamefont {Gisin}},
  \bibinfo {author} {\bibfnamefont {V.}~\bibnamefont {Scarani}}, \ and\
  \bibinfo {author} {\bibfnamefont {C.}~\bibnamefont {Simon}},\ }\href
  {\doibase 10.1103/PhysRevLett.98.220403} {\bibfield  {journal} {\bibinfo
  {journal} {Phys. Rev. Lett.}\ }\textbf {\bibinfo {volume} {98}},\ \bibinfo
  {pages} {220403} (\bibinfo {year} {2007})}\BibitemShut {NoStop}%
\bibitem [{\citenamefont {Larsson}(1998)}]{Larsson1998}%
  \BibitemOpen
  \bibfield  {author} {\bibinfo {author} {\bibfnamefont {J.-{\r{A}}.}\
  \bibnamefont {Larsson}},\ }\href {\doibase 10.1103/PhysRevA.57.3304}
  {\bibfield  {journal} {\bibinfo  {journal} {Phys. Rev. A}\ }\textbf {\bibinfo
  {volume} {57}},\ \bibinfo {pages} {3304} (\bibinfo {year}
  {1998})}\BibitemShut {NoStop}%
\bibitem [{\citenamefont {Larsson}\ and\ \citenamefont
  {Semitecolos}(2001)}]{Larsson2001}%
  \BibitemOpen
  \bibfield  {author} {\bibinfo {author} {\bibfnamefont {J.-{\r{A}}.}\
  \bibnamefont {Larsson}}\ and\ \bibinfo {author} {\bibfnamefont
  {J.}~\bibnamefont {Semitecolos}},\ }\href {\doibase
  10.1103/PhysRevA.63.022117} {\bibfield  {journal} {\bibinfo  {journal} {Phys.
  Rev. A}\ }\textbf {\bibinfo {volume} {63}},\ \bibinfo {pages} {022117}
  (\bibinfo {year} {2001})}\BibitemShut {NoStop}%
\bibitem [{\citenamefont {Garg}\ and\ \citenamefont {Mermin}(1987)}]{Garg1987}%
  \BibitemOpen
  \bibfield  {author} {\bibinfo {author} {\bibfnamefont {A.}~\bibnamefont
  {Garg}}\ and\ \bibinfo {author} {\bibfnamefont {N.~D.}\ \bibnamefont
  {Mermin}},\ }\href {\doibase 10.1103/PhysRevD.35.3831} {\bibfield  {journal}
  {\bibinfo  {journal} {Phys. Rev. D}\ }\textbf {\bibinfo {volume} {35}},\
  \bibinfo {pages} {3831} (\bibinfo {year} {1987})}\BibitemShut {NoStop}%
\bibitem [{\citenamefont {Pearle}(1970)}]{Pearle1970}%
  \BibitemOpen
  \bibfield  {author} {\bibinfo {author} {\bibfnamefont {P.~M.}\ \bibnamefont
  {Pearle}},\ }\href {\doibase 10.1103/PhysRevD.2.1418} {\bibfield  {journal}
  {\bibinfo  {journal} {Phys. Rev. D}\ }\textbf {\bibinfo {volume} {2}},\
  \bibinfo {pages} {1418} (\bibinfo {year} {1970})}\BibitemShut {NoStop}%
\bibitem [{\citenamefont {Howell}\ \emph {et~al.}(2002)\citenamefont {Howell},
  \citenamefont {Lamas-Linares},\ and\ \citenamefont
  {Bouwmeester}}]{Howell2002}%
  \BibitemOpen
  \bibfield  {author} {\bibinfo {author} {\bibfnamefont {J.}~\bibnamefont
  {Howell}}, \bibinfo {author} {\bibfnamefont {A.}~\bibnamefont
  {Lamas-Linares}}, \ and\ \bibinfo {author} {\bibfnamefont {D.}~\bibnamefont
  {Bouwmeester}},\ }\href {\doibase 10.1103/PhysRevLett.88.030401} {\bibfield
  {journal} {\bibinfo  {journal} {Phys. Rev. Lett.}\ }\textbf {\bibinfo
  {volume} {88}},\ \bibinfo {pages} {030401} (\bibinfo {year}
  {2002})}\BibitemShut {NoStop}%
\bibitem [{\citenamefont {Mermin}(1980)}]{Mermin1980}%
  \BibitemOpen
  \bibfield  {author} {\bibinfo {author} {\bibfnamefont {N.~D.}\ \bibnamefont
  {Mermin}},\ }\href {\doibase 10.1103/PhysRevD.22.356} {\bibfield  {journal}
  {\bibinfo  {journal} {Phys. Rev. D}\ }\textbf {\bibinfo {volume} {22}},\
  \bibinfo {pages} {356} (\bibinfo {year} {1980})}\BibitemShut {NoStop}%
\bibitem [{\citenamefont {Carter}\ \emph {et~al.}(1987)\citenamefont {Carter},
  \citenamefont {Drummond}, \citenamefont {Reid},\ and\ \citenamefont
  {Shelby}}]{Carter1987}%
  \BibitemOpen
  \bibfield  {author} {\bibinfo {author} {\bibfnamefont {S.~J.}\ \bibnamefont
  {Carter}}, \bibinfo {author} {\bibfnamefont {P.~D.}\ \bibnamefont
  {Drummond}}, \bibinfo {author} {\bibfnamefont {M.~D.}\ \bibnamefont {Reid}},
  \ and\ \bibinfo {author} {\bibfnamefont {R.~M.}\ \bibnamefont {Shelby}},\
  }\href {\doibase 10.1103/PhysRevLett.58.1841} {\bibfield  {journal} {\bibinfo
   {journal} {Phys. Rev. Lett.}\ }\textbf {\bibinfo {volume} {58}},\ \bibinfo
  {pages} {1841} (\bibinfo {year} {1987})}\BibitemShut {NoStop}%
\bibitem [{\citenamefont {Drummond}\ and\ \citenamefont
  {Carter}(1987)}]{Drummond1987}%
  \BibitemOpen
  \bibfield  {author} {\bibinfo {author} {\bibfnamefont {P.~D.}\ \bibnamefont
  {Drummond}}\ and\ \bibinfo {author} {\bibfnamefont {S.~J.}\ \bibnamefont
  {Carter}},\ }\href {\doibase 10.1364/JOSAB.4.001565} {\bibfield  {journal}
  {\bibinfo  {journal} {J. Opt. Soc. Am. B}\ }\textbf {\bibinfo {volume} {4}},\
  \bibinfo {pages} {1565} (\bibinfo {year} {1987})}\BibitemShut {NoStop}%
\bibitem [{\citenamefont {Werner}\ \emph {et~al.}(1995)\citenamefont {Werner},
  \citenamefont {Raymer}, \citenamefont {Beck},\ and\ \citenamefont
  {Drummond}}]{Werner1995}%
  \BibitemOpen
  \bibfield  {author} {\bibinfo {author} {\bibfnamefont {M.~J.}\ \bibnamefont
  {Werner}}, \bibinfo {author} {\bibfnamefont {M.~G.}\ \bibnamefont {Raymer}},
  \bibinfo {author} {\bibfnamefont {M.}~\bibnamefont {Beck}}, \ and\ \bibinfo
  {author} {\bibfnamefont {P.~D.}\ \bibnamefont {Drummond}},\ }\href {\doibase
  10.1103/PhysRevA.52.4202} {\bibfield  {journal} {\bibinfo  {journal} {Phys.
  Rev. A}\ }\textbf {\bibinfo {volume} {52}},\ \bibinfo {pages} {4202}
  (\bibinfo {year} {1995})}\BibitemShut {NoStop}%
\bibitem [{\citenamefont {Cahill}\ and\ \citenamefont
  {Glauber}(1969)}]{Cahill1969}%
  \BibitemOpen
  \bibfield  {author} {\bibinfo {author} {\bibfnamefont {K.~E.}\ \bibnamefont
  {Cahill}}\ and\ \bibinfo {author} {\bibfnamefont {R.~J.}\ \bibnamefont
  {Glauber}},\ }\href {\doibase 10.1103/PhysRev.177.1857} {\bibfield  {journal}
  {\bibinfo  {journal} {Phys. Rev.}\ }\textbf {\bibinfo {volume} {177}},\
  \bibinfo {pages} {1857} (\bibinfo {year} {1969})}\BibitemShut {NoStop}%
\bibitem [{\citenamefont {Corney}\ and\ \citenamefont
  {Drummond}(2003)}]{Corney2003}%
  \BibitemOpen
  \bibfield  {author} {\bibinfo {author} {\bibfnamefont {J.~F.}\ \bibnamefont
  {Corney}}\ and\ \bibinfo {author} {\bibfnamefont {P.~D.}\ \bibnamefont
  {Drummond}},\ }\href {\doibase 10.1103/PhysRevA.68.063822} {\bibfield
  {journal} {\bibinfo  {journal} {Phys. Rev. A}\ }\textbf {\bibinfo {volume}
  {68}},\ \bibinfo {pages} {063822} (\bibinfo {year} {2003})}\BibitemShut
  {NoStop}%
\bibitem [{\citenamefont {Corney}\ and\ \citenamefont
  {Drummond}(2004)}]{Corney2004}%
  \BibitemOpen
  \bibfield  {author} {\bibinfo {author} {\bibfnamefont {J.~F.}\ \bibnamefont
  {Corney}}\ and\ \bibinfo {author} {\bibfnamefont {P.~D.}\ \bibnamefont
  {Drummond}},\ }\href {\doibase 10.1103/PhysRevLett.93.260401} {\bibfield
  {journal} {\bibinfo  {journal} {Phys. Rev. Lett.}\ }\textbf {\bibinfo
  {volume} {93}},\ \bibinfo {pages} {260401} (\bibinfo {year}
  {2004})}\BibitemShut {NoStop}%
\bibitem [{\citenamefont {Glauber}(1963)}]{Glauber1963-states}%
  \BibitemOpen
  \bibfield  {author} {\bibinfo {author} {\bibfnamefont {R.~J.}\ \bibnamefont
  {Glauber}},\ }\href {\doibase 10.1103/PhysRev.131.2766} {\bibfield  {journal}
  {\bibinfo  {journal} {Phys. Rev.}\ }\textbf {\bibinfo {volume} {131}},\
  \bibinfo {pages} {2766} (\bibinfo {year} {1963})}\BibitemShut {NoStop}%
\bibitem [{\citenamefont {Sudarshan}(1963)}]{Sudarshan1963}%
  \BibitemOpen
  \bibfield  {author} {\bibinfo {author} {\bibfnamefont {E.}~\bibnamefont
  {Sudarshan}},\ }\href {\doibase 10.1103/PhysRevLett.10.277} {\bibfield
  {journal} {\bibinfo  {journal} {Phys. Rev. Lett.}\ }\textbf {\bibinfo
  {volume} {10}},\ \bibinfo {pages} {277} (\bibinfo {year} {1963})}\BibitemShut
  {NoStop}%
\bibitem [{\citenamefont {Gupta}\ and\ \citenamefont {Song}(1997)}]{Gupta1997}%
  \BibitemOpen
  \bibfield  {author} {\bibinfo {author} {\bibfnamefont {A.~K.}\ \bibnamefont
  {Gupta}}\ and\ \bibinfo {author} {\bibfnamefont {D.}~\bibnamefont {Song}},\
  }\href {\doibase 10.1016/S0378-3758(96)00129-2} {\bibfield  {journal}
  {\bibinfo  {journal} {J. Stat. Plan. Infer.}\ }\textbf {\bibinfo {volume}
  {60}},\ \bibinfo {pages} {241} (\bibinfo {year} {1997})}\BibitemShut
  {NoStop}%
\bibitem [{\citenamefont {Marsaglia}(1972)}]{Marsaglia1972}%
  \BibitemOpen
  \bibfield  {author} {\bibinfo {author} {\bibfnamefont {G.}~\bibnamefont
  {Marsaglia}},\ }\href {\doibase 10.1214/aoms/1177692644} {\bibfield
  {journal} {\bibinfo  {journal} {Ann. Math. Statist.}\ }\textbf {\bibinfo
  {volume} {43}},\ \bibinfo {pages} {645} (\bibinfo {year} {1972})}\BibitemShut
  {NoStop}%
\bibitem [{\citenamefont {Muller}(1959)}]{Muller1959}%
  \BibitemOpen
  \bibfield  {author} {\bibinfo {author} {\bibfnamefont {M.~E.}\ \bibnamefont
  {Muller}},\ }\href {\doibase 10.1145/377939.377946} {\bibfield  {journal}
  {\bibinfo  {journal} {Commun. ACM}\ }\textbf {\bibinfo {volume} {2}},\
  \bibinfo {pages} {19} (\bibinfo {year} {1959})}\BibitemShut {NoStop}%
\bibitem [{\citenamefont {Rosales-Z\'{a}rate}\ and\ \citenamefont
  {Drummond}(2011)}]{Rosales-Zarate2011}%
  \BibitemOpen
  \bibfield  {author} {\bibinfo {author} {\bibfnamefont {L.~E.~C.}\
  \bibnamefont {Rosales-Z\'{a}rate}}\ and\ \bibinfo {author} {\bibfnamefont
  {P.~D.}\ \bibnamefont {Drummond}},\ }\href {\doibase
  10.1103/PhysRevA.84.042114} {\bibfield  {journal} {\bibinfo  {journal} {Phys.
  Rev. A}\ }\textbf {\bibinfo {volume} {84}},\ \bibinfo {pages} {042114}
  (\bibinfo {year} {2011})}\BibitemShut {NoStop}%
\bibitem [{\citenamefont {Drummond}\ and\ \citenamefont
  {Mortimer}(1991)}]{Drummond1991}%
  \BibitemOpen
  \bibfield  {author} {\bibinfo {author} {\bibfnamefont {P.~D.}\ \bibnamefont
  {Drummond}}\ and\ \bibinfo {author} {\bibfnamefont {I.~K.}\ \bibnamefont
  {Mortimer}},\ }\href {\doibase 10.1016/0021-9991(91)90077-X} {\bibfield
  {journal} {\bibinfo  {journal} {J. Comput. Phys.}\ }\textbf {\bibinfo
  {volume} {93}},\ \bibinfo {pages} {144} (\bibinfo {year} {1991})}\BibitemShut
  {NoStop}%
\bibitem [{\citenamefont {Hillery}\ and\ \citenamefont
  {Zubairy}(2006)}]{Hillery2006}%
  \BibitemOpen
  \bibfield  {author} {\bibinfo {author} {\bibfnamefont {M.}~\bibnamefont
  {Hillery}}\ and\ \bibinfo {author} {\bibfnamefont {M.~S.}\ \bibnamefont
  {Zubairy}},\ }\href {\doibase 10.1103/PhysRevLett.96.050503} {\bibfield
  {journal} {\bibinfo  {journal} {Phys. Rev. Lett.}\ }\textbf {\bibinfo
  {volume} {96}},\ \bibinfo {pages} {050503} (\bibinfo {year}
  {2006})}\BibitemShut {NoStop}%
\bibitem [{\citenamefont {Raymer}\ \emph {et~al.}(1991)\citenamefont {Raymer},
  \citenamefont {Drummond},\ and\ \citenamefont {Carter}}]{Raymer1991}%
  \BibitemOpen
  \bibfield  {author} {\bibinfo {author} {\bibfnamefont {M.~G.}\ \bibnamefont
  {Raymer}}, \bibinfo {author} {\bibfnamefont {P.~D.}\ \bibnamefont
  {Drummond}}, \ and\ \bibinfo {author} {\bibfnamefont {S.~J.}\ \bibnamefont
  {Carter}},\ }\href {\doibase 10.1364/OL.16.001189} {\bibfield  {journal}
  {\bibinfo  {journal} {Opt. Lett.}\ }\textbf {\bibinfo {volume} {16}},\
  \bibinfo {pages} {1189} (\bibinfo {year} {1991})}\BibitemShut {NoStop}%
\bibitem [{\citenamefont {Werner}\ and\ \citenamefont
  {Drummond}(1997)}]{Werner1997}%
  \BibitemOpen
  \bibfield  {author} {\bibinfo {author} {\bibfnamefont {M.~J.}\ \bibnamefont
  {Werner}}\ and\ \bibinfo {author} {\bibfnamefont {P.~D.}\ \bibnamefont
  {Drummond}},\ }\href {\doibase 10.1006/jcph.1996.5638} {\bibfield  {journal}
  {\bibinfo  {journal} {J. Comput. Phys.}\ }\textbf {\bibinfo {volume} {132}},\
  \bibinfo {pages} {312} (\bibinfo {year} {1997})}\BibitemShut {NoStop}%
\bibitem [{\citenamefont {U'Ren}\ \emph {et~al.}(2004)\citenamefont {U'Ren},
  \citenamefont {Silberhorn}, \citenamefont {Banaszek},\ and\ \citenamefont
  {Walmsley}}]{U'Ren2004}%
  \BibitemOpen
  \bibfield  {author} {\bibinfo {author} {\bibfnamefont {A.~B.}\ \bibnamefont
  {U'Ren}}, \bibinfo {author} {\bibfnamefont {C.}~\bibnamefont {Silberhorn}},
  \bibinfo {author} {\bibfnamefont {K.}~\bibnamefont {Banaszek}}, \ and\
  \bibinfo {author} {\bibfnamefont {I.~A.}\ \bibnamefont {Walmsley}},\ }\href
  {\doibase 10.1103/PhysRevLett.93.093601} {\bibfield  {journal} {\bibinfo
  {journal} {Phys. Rev. Lett.}\ }\textbf {\bibinfo {volume} {93}},\ \bibinfo
  {pages} {093601} (\bibinfo {year} {2004})}\BibitemShut {NoStop}%
\bibitem [{\citenamefont {Fedrizzi}\ \emph {et~al.}(2007)\citenamefont
  {Fedrizzi}, \citenamefont {Herbst}, \citenamefont {Poppe}, \citenamefont
  {Jennewein},\ and\ \citenamefont {Zeilinger}}]{Fedrizzi2007}%
  \BibitemOpen
  \bibfield  {author} {\bibinfo {author} {\bibfnamefont {A.}~\bibnamefont
  {Fedrizzi}}, \bibinfo {author} {\bibfnamefont {T.}~\bibnamefont {Herbst}},
  \bibinfo {author} {\bibfnamefont {A.}~\bibnamefont {Poppe}}, \bibinfo
  {author} {\bibfnamefont {T.}~\bibnamefont {Jennewein}}, \ and\ \bibinfo
  {author} {\bibfnamefont {A.}~\bibnamefont {Zeilinger}},\ }\href {\doibase
  10.1364/OE.15.015377} {\bibfield  {journal} {\bibinfo  {journal} {Opt.
  Express}\ }\textbf {\bibinfo {volume} {15}},\ \bibinfo {pages} {15377}
  (\bibinfo {year} {2007})}\BibitemShut {NoStop}%
\bibitem [{\citenamefont {Reid}\ \emph {et~al.}(2014)\citenamefont {Reid},
  \citenamefont {Opanchuk}, \citenamefont {Rosales-Z\'{a}rate},\ and\
  \citenamefont {Drummond}}]{Reid2014-60qubit}%
  \BibitemOpen
  \bibfield  {author} {\bibinfo {author} {\bibfnamefont {M.~D.}\ \bibnamefont
  {Reid}}, \bibinfo {author} {\bibfnamefont {B.}~\bibnamefont {Opanchuk}},
  \bibinfo {author} {\bibfnamefont {L.}~\bibnamefont {Rosales-Z\'{a}rate}}, \
  and\ \bibinfo {author} {\bibfnamefont {P.~D.}\ \bibnamefont {Drummond}},\
  }\href {http://arxiv.org/abs/1406.2432v1} {\  (\bibinfo {year} {2014})},\
  \Eprint {http://arxiv.org/abs/1406.2432} {arXiv:1406.2432} \BibitemShut
  {NoStop}%
\bibitem [{\citenamefont {Rohde}\ \emph {et~al.}(2014)\citenamefont {Rohde},
  \citenamefont {Motes}, \citenamefont {Knott},\ and\ \citenamefont
  {Munro}}]{Rohde2014}%
  \BibitemOpen
  \bibfield  {author} {\bibinfo {author} {\bibfnamefont {P.~P.}\ \bibnamefont
  {Rohde}}, \bibinfo {author} {\bibfnamefont {K.~R.}\ \bibnamefont {Motes}},
  \bibinfo {author} {\bibfnamefont {P.~A.}\ \bibnamefont {Knott}}, \ and\
  \bibinfo {author} {\bibfnamefont {W.~J.}\ \bibnamefont {Munro}},\ }\href
  {http://arxiv.org/abs/1401.2199} {\  (\bibinfo {year} {2014})},\ \Eprint
  {http://arxiv.org/abs/1401.2199} {arXiv:1401.2199} \BibitemShut {NoStop}%
\end{thebibliography}%

\end{document}